\documentclass[pre,aps,amsmath,amssymb,amsfonts,floatfix,showpacs,twocolumn,footinbib]{revtex4-1}

\usepackage{graphicx}
\usepackage{amsmath,amsfonts}
\usepackage{xcolor}
\usepackage{color}
\usepackage{mathtools}
\usepackage{eufrak}
\usepackage{pgfplots}

\newcommand{\up}{\uparrow}
\newcommand{\dn}{\downarrow}
\newcommand{\kv}{\ensuremath{\mathbf{k}}}
\newcommand{\qv}{\ensuremath{\mathbf{q}}}
\newcommand{\Qv}{\ensuremath{\mathbf{Q}}}

\newcommand{\ch}{\ensuremath{\text{ch}}}
\newcommand{\sz}{\ensuremath{\text{sp}}}
\newcommand{\av}[1]{\ensuremath{\left\langle #1 \right\rangle}}

\usepackage{tikz}

\usetikzlibrary{calc}
\usetikzlibrary{decorations.pathmorphing}
\usetikzlibrary{decorations.pathreplacing}
\usetikzlibrary{decorations.markings}
\usetikzlibrary{shapes.misc}
\usetikzlibrary{shapes}
\usetikzlibrary{positioning}
\usetikzlibrary{snakes}
\usetikzlibrary{arrows}
\usetikzlibrary{fadings}
\usetikzlibrary{calc,arrows}

\tikzset{snake it/.style={decorate, decoration=snake}}
\usetikzlibrary{3d}
\usepackage{mathtools}
    \tikzset{
            partial ellipse/.style args={#1:#2:#3}{
                        insert path={+ (#1:#3) arc (#1:#2:#3)}
                            }
                        }

\usetikzlibrary{calc}
\tikzset{
            inertial frame/.style = {x={(-20:2cm)}, y={(-160:2cm)}, z={(90:2cm)}},
              local frame/.style = {shift={(local origin)}, x={(40:.7cm)}, y={(150:.7cm)}, z={(105:.7cm)}}
          }

    \tikzset{middlearrow/.style={
                decoration={markings,
                            mark= at position 0.5 with {\arrow{#1}} ,
                                    },
                                            postaction={decorate}
                                                }
                                                }
\tikzset{cross/.style={cross out, draw, 
         minimum size=2*(#1-\pgflinewidth), 
                  inner sep=0pt, outer sep=0pt}}

\def\presuper#1#2%
  {\mathop{}%
   \mathopen{\vphantom{#2}}^{#1}%
   \kern-0.5\scriptspace%
   #2}

\usepackage{hyperref}
\hypersetup{colorlinks=true,breaklinks,linkcolor=blue,urlcolor=blue,citecolor=blue}
\begin{document}

    \pgfmathdeclarefunction{gauss}{2}{%
          \pgfmathparse{1/(#2*sqrt(2*pi))*exp(-((x-#1)^2)/(2*#2^2))}%
          }
    \pgfmathdeclarefunction{mgauss}{2}{%
          \pgfmathparse{-1/(#2*sqrt(2*pi))*exp(-((x-#1)^2)/(2*#2^2))}%
          }
    \pgfmathdeclarefunction{lorentzian}{2}{%
        \pgfmathparse{1/(#2*pi)*((#2)^2)/((x-#1)^2+(#2)^2)}%
          }
    \pgfmathdeclarefunction{mlorentzian}{2}{%
        \pgfmathparse{-1/(#2*pi)*((#2)^2)/((x-#1)^2+(#2)^2)}%
          }

\author{Friedrich Krien}
\affiliation{International School for Advanced Studies, SISSA, Trieste, Italy}

\title{Efficient evaluation of the polarization function in the dynamical mean-field theory}

\begin{abstract}
The dynamical susceptibility of strongly correlated electronic systems can be calculated within the framework of the dynamical mean-field theory (DMFT).
The required measurement of the four-point vertex of the auxiliary impurity model is however costly and restricted to a finite grid of Matsubara frequencies,
leading to a cutoff error.
It is shown that the propagation of this error to the lattice response function can be minimized by virtue of an exact
decomposition of the DMFT polarization function into local and nonlocal parts. The former is measured directly by the impurity solver,
while the latter is given in terms of a ladder equation for the Hedin vertex that features an unprecedentedly
fast decay of frequency summations compared to previous calculation schemes, such as the one of the dual boson approach.
At strong coupling the local approximation of the TRILEX approach is viable,
but vertex corrections to the polarization should be dropped on equal footing to recover the correct prefactor of the effective exchange.
In finite dimensions the DMFT susceptibility exhibits spurious mean-field criticality, therefore, a two-particle self-consistent and frequency-dependent
correction term is introduced, similar to the Moriya-$\lambda$ correction of the dynamical vertex approximation.
Applications to the two- and three-dimensional Hubbard models on the square and cubic lattices show that the
expected critical behavior near an antiferromagnetic instability is recovered.
\end{abstract}

\maketitle

\section{Introduction}
The dynamical mean-field theory (DMFT) is a powerful non-perturbative approach to strong local correlations in the Hubbard model~\cite{Georges96}.
Although in widespread use, many aspects of the DMFT are still under investigation,
which is fueled to large extent by persistent algorithmic advances in the solution of its auxiliary Anderson impurity model~\cite{Gull11}.
These improvements allow insights into the two-particle level of the DMFT approximation~\cite{Rohringer12},
which is also the elemental precursor for its diagrammatic extensions~\cite{Rohringer17}.

A basic application for DMFT at the two-particle level is the calculation of the dynamical susceptibility,
which allows to study, for example, phase transitions~\cite{Georges96}, the electron energy loss spectrum~\cite{vanLoon14-2},
nuclear relaxation rate~\cite{Boehnke12}, and Goldstone excitations~\cite{Geffroy18}.
The DMFT susceptibility is furthermore an integral part of the ladder dynamical vertex approximation~\cite{Toschi07,Galler17}.
Calculation of this correlation function however requires knowledge of the impurity vertex function,
which is often evaluated by means of improved estimators for continuous-time quantum Monte-Carlo (CTQMC) solvers~\cite{Hafermann12,Gunacker16}.
The further development of improved estimators is highly desirable, as they allow to efficiently calculate the DMFT susceptibility in multi-orbital settings,
see for example Refs.~\cite{Boehnke12,Boehnke18,Geffroy18}.
Recently, progress has been reported in the measurement of the vertex function within the exact diagonalization method~\cite{Tanaka18}.

The role of the improved estimators in CTQMC methods is to minimize the statistical noise of the Monte Carlo measurement,
which for fixed run-time greatly increases with the number of dynamic degrees of freedom (Matsubara frequencies) of the measured quantity.
A further numerical error is introduced because the measurement of the impurity vertex function
is restricted to a finite grid of Matsubara frequencies. In order to obtain a gauge invariant lattice response function in DMFT
it is necessary to account for an infinite number, that is, a ladder of vertex corrections~\cite{Hafermann14-2}.
For each vertex correction the value of the impurity vertex at all frequencies enters the calculation,
and therefore due to the finite Matsubara grid a cutoff error arises. Consequently, the numerical error of the DMFT response function
may not only be minimized by an improved Monte Carlo measurement but also by reduction of the cutoff error.
A straightforward way to do this is to account for the asymptotics of the vertex function~\cite{Kunes11,Wentzell16,Kaufmann17,Tagliavini18}.

A further option for improvement, the subject of this work, is to use the numerically exact impurity solver to sum local diagrams exactly.
For concreteness, within the dual boson approach and in a calculation scheme by Pruschke et al. the DMFT susceptibility $X$
is written as the sum of local and nonlocal parts~\cite{Pruschke96,Rubtsov12,Hafermann14-2},
\begin{align}
    X_{\qv}(\omega)=\chi(\omega)+\tilde{X}_\qv(\omega)\label{eq:suscdb},
\end{align}
where $\qv$ is the lattice momentum and $\omega$ the (bosonic) Matsubara frequency.
The local part, the impurity susceptibility $\chi$, depends only on one frequency and is calculated directly by the impurity solver,
which in effect sums all local two-particle diagrams that taken together yield $\chi$.
Moreover, even though DMFT neglects nonlocal correlations, the lattice susceptibility takes local vertex corrections at \textit{different} lattice sites into account,
which give rise to the nonlocal term $\tilde{X}$ (see also Fig. 1 of Ref.~\cite{vanLoon16}).
The dual boson formula~\eqref{eq:suscdb} is numerically efficient because the statistical and cutoff errors
attached to the impurity vertex only affect the nonlocal term, not $\chi$.
Even when the asymptotic behavior of the impurity vertex is neglected it allows the analytical continuation 
of the susceptibility to the real axis~\cite{Hafermann14-2,vanLoon14-2}.
It is however desirable to preserve numerical resources, hence further improvements are welcome.

In this work it will be shown that the concept of breaking down the DMFT susceptibility into simpler diagrammatic pieces can be taken to a further level
by separating exactly the diagrams from the vertex function that are irreducible with respect to the bare Hubbard interaction $U$.
This leads to a decomposition of the polarization function $\Pi$, which is $U$-irreducible, into local and nonlocal parts~\cite{Stepanov16-2},
$\Pi_\qv(\omega)=\pi(\omega)+\tilde{\Pi}_\qv(\omega)$, analogous to equation~\eqref{eq:suscdb}.
The nonlocal part $\tilde{\Pi}$ is obtained via an efficient ladder equation for the Hedin three-leg vertex~\cite{Hedin65}.
The lattice polarization $\Pi_\qv(\omega)$ in turn encapsulates all non-trivial information about the two-particle spectrum.

The Hedin vertex also plays a central role in the TRILEX approach~\cite{Ayral15-2}.
In this method nonlocal vertex corrections to the Hedin vertex are neglected,
and therefore the calculation of the four-point vertex function of the impurity model is not necessary.
However, this approximation can be introduced in different ways, for example,
within the dual boson formalism it accounts for more vertex corrections than in TRILEX~\cite{Stepanov16-2}.
It is shown in this work that for large interaction these additional vertex corrections decide about the prefactor of the effective exchange coupling.
Both TRILEX and dual boson account for a nonlocal self-energy, however, this work focuses on approximations to the polarization function.

Lastly, a further aspect is considered in the application of the efficient formula for the polarization:
In finite dimensions the DMFT susceptibility violates the Pauli principle and suffers from a spurious mean-field instability in two dimensions.
It has been shown previously that the Mermin-Wagner theorem is satisfied in the renormalized ladder dual fermion approach~\cite{Otsuki14}
or after introduction of the Moriya-$\lambda$ correction to the DMFT susceptibility~\cite{Katanin09}.
In three dimensions both approaches renormalize the criticality of the underlying dynamical mean-field starting point~\cite{Rohringer11,Hirschmeier15}.
Furthermore, in the ladder dynamical vertex approximation the satisfaction of local charge and spin sum rules by the Moriya-$\lambda$ correction
is crucial to ensure the proper asymptotic behavior of the electronic self-energy~\cite{Katanin09,Rohringer16}.
Similar to the Moriya-$\lambda$ and two-particle self-consistent approach (TPSC)~\cite{Vilk97}, in this work the mean-field artifacts of the DMFT susceptibility
are removed by virtue of a frequency-dependent correction term that is fixed by a two-particle self-consistent constraint.
It is shown that this approach satisfies the Mermin-Wagner theorem and predicts the same criticality of the half-filled three-dimensional Hubbard model as the Moriya-$\lambda$.

The paper is organized as follows: The Hubbard Hamiltonian, the DMFT approximation, and the Anderson impurity model are briefly recollected in Sec.~\ref{sec:hubbard}.
The reducible and irreducible vertices of the impurity model are defined in Sec.~\ref{sec:impvertices}.
The efficient formula for the DMFT polarization is presented in Sec.~\ref{sec:pilattice} and compared to the dual boson formula.
A two-particle self-consistent modification of the DMFT susceptibility and a TRILEX-like approximation
are introduced in Sec.~\ref{sec:tpscdmf} and applied in Sec.~\ref{sec:results}. The conclusions follow in Sec.~\ref{sec:conclusions}.
A self-contained derivation of the ladder equation for the Hedin vertex is provided in the Appendices~\ref{app:irr}-\ref{app:hedin}.

\section{Hubbard Hamiltonian and DMFT approximation}\label{sec:hubbard}
The Hamiltonian of the paramagnetic two- or three-dimensional Hubbard model is given as,
\begin{align}
    H = &-\sum_{\langle ij\rangle\sigma}{t}_{ij} c^\dagger_{i\sigma}c^{}_{j\sigma}+ U\sum_{i} n_{i\up} n_{i\dn},\label{eq:hubbard}
\end{align}
where ${t}_{ij}$ is the nearest neighbor hopping between lattice sites $i,j$, its absolute value ${t}=1$ is the unit of energy.
$c^{},c^\dagger$ are the construction operators, $\sigma=\up,\dn$ the spin index.
$U$ is the Hubbard repulsion between the densities $n_{\sigma}=c^\dagger_{\sigma}c^{}_{\sigma}$. 

In the DMFT approximation the self-energy $\Sigma$ of Green's function is local,
\begin{align}
    G_k=[\imath \nu-\varepsilon_\kv+\mu-\Sigma(\nu)]^{-1}\label{eq:gf},
\end{align}
where $k=(\kv,\nu)$ comprises lattice momentum and fermionic Matsubara frequency, $\varepsilon_\kv$ is the dispersion, $\mu$ is the chemical potential.
$\Sigma(\nu)$ is the self-energy of an auxiliary Anderson impurity model (AIM) that is solved numerically exactly. The action of the AIM reads,
\begin{align}
  S_{\text{AIM}}=&-\sum_{\nu\sigma}c^*_{\nu\sigma}(\imath\nu+\mu-\Delta_\nu)c^{}_{\nu\sigma}+U \sum_\omega n_{\up\omega} n_{\dn\omega}.
  \label{eq:aim}
\end{align}
Here $\Delta_{\nu}$ denotes the hybridization function, $\omega$ is a bosonic Matsubara frequency,
summations $\sum_\nu,\sum_\omega$ imply multiplication with the temperature $T$. $c^*,c$ are Grassmann numbers.
In DMFT the hybridization function is fixed self-consistently according to the constraint,
\begin{align}
    \sum_\kv G_k=g_\nu,\label{eq:selfconsistency}
\end{align}
where $g$ is the numerically exact local Green's function of the AIM. 
Note that summation over $\kv$ implies division by the number of lattice sites $N$.

\section{Impurity vertices}\label{sec:impvertices}
The calculation of the dynamical susceptibility requires knowledge of higher correlation functions of the impurity.
Directly measured by the solver are the susceptibility,
$\chi^\alpha_\omega=-\langle{\rho^\alpha_{-\omega}\rho^\alpha_\omega}\rangle+\beta\langle n\rangle\langle n\rangle\delta_\omega\delta_{\alpha,\ch}$, the four-point, 
\begin{align}
    g^{(4),\alpha}_{\nu\nu'\omega}=&-\frac{1}{2}\sum_{\sigma_i}s^\alpha_{\sigma_1'\sigma_1}s^\alpha_{\sigma_2'\sigma_2}
    \langle{c^{}_{\nu\sigma_1}c^{*}_{\nu+\omega,\sigma_1'}c^{}_{\nu'+\omega,\sigma_2}c^{*}_{\nu'\sigma_2'}}\rangle\notag,
\end{align}
and the three-point function,
\begin{align}
    g^{(3),\alpha}_{\nu\omega}=&\frac{1}{2}\sum_{\sigma\sigma'}s^\alpha_{\sigma'\sigma}
    \langle{c^{}_{\nu\sigma}c^{*}_{\nu+\omega,\sigma'}\rho^\alpha_\omega}\rangle\notag=\sum_{\nu'}g^{(4),\alpha}_{\nu\nu'\omega},
\end{align}
where $s^\alpha$ are the Pauli matrices ($\alpha=\ch,\sz$), $\rho^\ch=n_\up+n_\dn$ and $\rho^\sz=n_\up-n_\dn$ are the charge and spin densities.

\subsection{Reducible vertices}
One defines the four- and three-point vertices $f$ and $\bar{\lambda}$,
\begin{align}
f^\alpha_{\nu\nu'\omega}=&\frac{g^{(4),\alpha}_{\nu\nu'\omega}-\beta g_\nu g_{\nu+\omega}\delta_{\nu\nu'}+2\beta g_\nu g_{\nu'}\delta_{\omega}\delta_{\alpha,\ch}}
    {g_\nu g_{\nu+\omega}g_{\nu'}g_{\nu'+\omega}}\label{eq:4pvertex},\\
\bar{\lambda}^\alpha_{\nu\omega}=&\frac{g^{(3),\alpha}_{\nu\omega}+\beta g_\nu \langle n\rangle\delta_{\omega}\delta_{\alpha,\ch}}{g_\nu g_{\nu+\omega}}.\label{eq:3pvertex}
\end{align}
Although numerically unfavorable $\bar{\lambda}$ can in principle also be obtained by attaching legs to $f$ from the \text{right} and adding $1$,
$\bar{\lambda}^\alpha_{\nu\omega}=1+\sum_{\nu'}f^\alpha_{\nu\nu'\omega}g_{\nu'}g_{\nu'+\omega}$,
therefore $\bar{\lambda}$ is a \textit{right-sided} three-leg vertex~\cite{vanLoon18}, the \textit{left-sided} one $\lambda$ is obtained by attaching the legs from the left
or via the symmetry relation, $\bar{\lambda}^\alpha_{\nu\omega}=\lambda^\alpha_{\nu+\omega,-\omega}$.

\begin{figure}
\begin{center}
  \includegraphics[width=0.47\textwidth]{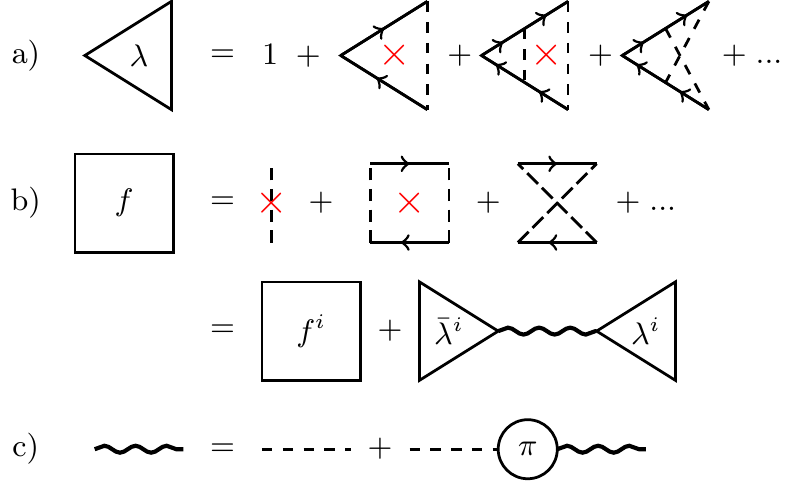}
\end{center}
    \caption{\label{fig:irrvertices}
    Lowest order contributions to the $U^\alpha$-\textit{reducible} three-leg [a)] and four-leg [b)] vertices $\lambda$ and $f$ of the impurity.
    Dashed lines indicate the bare interaction $\pm U$, arrows the {impurity} Green's function $g$. 
    Red crosses mark RPA-like contributions that are \textit{not} included in the $U^\alpha$-\textit{irreducible} $\lambda^i$ and $f^i$.
    The second line of diagram b) shows the relation between $f, f^i$, and $\lambda^i$,
    the wiggly line denotes the screened interaction $w$ of the impurity [cf. Eq.~\eqref{eq:firrtext} and Appendix~\ref{app:irr}],
    which is represented in diagram c) as a geometric series.
    }
    \end{figure}

\subsection{$U^\alpha$-irreducible vertices}\label{sec:hedin}
In order to make the later calculation of the DMFT lattice correlation functions efficient
the impurity vertices are decomposed following Hertz and Edwards~\cite{Hertz73}:

The diagram $a)$ in Fig.~\ref{fig:irrvertices} shows that when the three-leg vertex $\lambda^\alpha$ is expanded diagrammatically
one may encounter, in going from left to right, an insertion of the bare interaction $U^\alpha$, where $U^\ch=+U$ or $U^\sz=-U$.
The Hubbard interaction is just a constant, and the incoming impurity Green's function lines on the left of $U^\alpha$ can thus be contracted,
the same is case for the out-going lines.

On the left of $U^\alpha$ there hence arises a contribution to the $U^\alpha$-\textit{irreducible} polarization $\pi^\alpha$
of the impurity [related to the susceptibility via $\chi^\alpha_\omega=2\pi^\alpha_\omega/(1-U^\alpha\pi^\alpha)$],
whereas on the right of $U^\alpha$ begins once again an expansion of the three-leg vertex.
As shown algebraically in Appendix~\ref{app:irr}, one thus separates diagrams from $\lambda$ that are once or manifold $U^\alpha$-\textit{reducible},
\begin{align}
    \lambda^\alpha_{\nu\omega}=&\frac{\lambda^{i,\alpha}_{\nu\omega}}{1-U^{\alpha}\pi^\alpha_\omega},\label{eq:lambdairrtext}
\end{align}
where $\lambda^i$ is the $U^\alpha$-\textit{irreducible} three-leg vertex -- the Hedin vertex -- of the impurity.

Let us perform this procedure also for the four-point vertex $f$, as depicted in Fig.~\ref{fig:irrvertices} b).
$f$ obviously contains one part $f^i$ that is irreducible, whereas in the remaining terms one finds at least one insertion $U^\alpha$.
At this point the incoming lines may be closed and a \textit{right-sided} Hedin vertex $\bar{\lambda}^i$ arises on the left of $U^\alpha$.
In fact, also on the right of $U^\alpha$ the lines may be closed, which means that a true four-point contribution does not arise in the
$U^\alpha$-\textit{reducible} diagrams. For this reason the whole of the reducible diagrams may be split into the three- and two-point objects
$\bar{\lambda}^i,\lambda^i$ and $\pi$, respectively,
\begin{align}
  f^\alpha_{\nu\nu'\omega}=&f^{i,\alpha}_{\nu\nu'\omega}+\bar{\lambda}^{i,\alpha}_{\nu\omega}\,w^\alpha_\omega\,\lambda^{i,\alpha}_{\nu'\omega},\label{eq:firrtext}
\end{align}
where $w^\alpha_\omega={U^{\alpha}}/({1-U^\alpha\pi^\alpha_{q}})$ is the screened interaction of the impurity [cf. Fig.~\ref{fig:irrvertices} c)].

The equations~\eqref{eq:lambdairrtext} and~\eqref{eq:firrtext} are valuable because they separate RPA-like diagrams from the vertices $\lambda$ and $f$,
which are absorbed into the geometric series in Fig.~\ref{fig:irrvertices} c), the screened interaction $w$~\footnote{
    The polarization $\pi$ can indeed be interpreted as the `self-energy' of the screened interaction $w$,
    analogous to the Dyson equation $g=g^0/(1-g^0\Sigma)$, and the bare interaction $U$ assumes the role of the bare Green's function $g^0$.
    On the other hand, $U$ also corresponds to the \textit{two-particle} self-energy of the RPA approximation~\cite{Mahan00,Vilk96},
    one may therefore refer to the diagrams in Fig.~\ref{fig:irrvertices} c) as `RPA-like'.
}.
Similar relations are also valid for the Hubbard model~\eqref{eq:hubbard}, see Ref.~\cite{Held11} and Appendix~\ref{app:irr}.
The characteristic triangle-wiggle-triangle diagram in Fig.~\ref{fig:irrvertices} b)
is typically large when the corresponding susceptibility $\chi^\alpha$ is large, since then $U^\alpha\pi^\alpha\approx1$.

One should note that in the reducible contribution $\bar{\lambda}^{i}_{\nu\omega}w_\omega\lambda^{i}_{\nu'\omega}$ in Eq.~\eqref{eq:firrtext} the dependence on $\nu$ and $\nu'$ is separated.
Therefore, this term is comprised in the lowest order of a singular value decomposition of $f$~\cite{Otsuki19}.

\section{Efficient formula}\label{sec:pilattice}
\begin{figure}
    \begin{center}
      \includegraphics[width=0.45\textwidth]{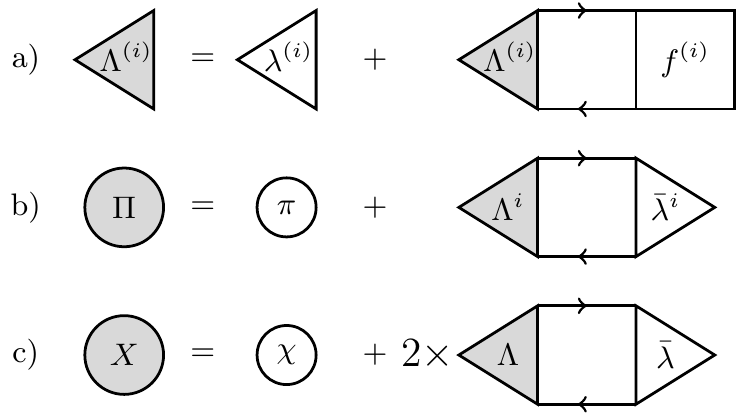}
    \end{center}
    \caption{\label{fig:diagrams}
    a) DMFT approximation to the three-leg vertices $\Lambda$ and $\Lambda^i$ (full triangles).
    Bare triangle and box represent the three- and four-leg vertices of the impurity. Arrows denote nonlocal Green's functions $\tilde{G}$.
    b) The lattice polarization $\Pi$ (full circle) is given as the sum of the impurity polarization $\pi$ (bare circle) and nonlocal corrections.
    Note that the latter are given by two Hedin vertices on the left and right, whereas in the original TRILEX there is only one~\cite{Ayral15-2}.
    c) Dual boson formula~\eqref{eq:suscdb} for the susceptibility. In this case the nonlocal corrections are given by the $U$-\textit{reducible} three-leg vertices $\Lambda,\lambda$.
    }
    \end{figure}

The goal is to calculate the dynamical susceptibility in the DMFT approximation, see also definition~\eqref{app:xdef},
\begin{align}
    X^\alpha_q=\frac{2\Pi^\alpha_q}{1-U^\alpha\Pi^\alpha_q},\label{eq:susctext}
\end{align}
where $q=(\qv,\omega)$ and $\Pi^\alpha_q$ is the lattice polarization.
The form of equation~\eqref{eq:susctext} resembles the RPA susceptibility, however,
the polarization $\Pi$ is similar to the Lindhard function only in the weak coupling limit,
while for intermediate and large coupling $\Pi$ is strongly renormalized by the frequency
dependence of the DMFT self-energy $\Sigma(\nu)$ and vertex corrections~\cite{Hafermann14-2}.
The latter can be taken into account in the following efficient way:

It is shown in Appendix~\ref{app:susc} that in DMFT the polarization can be decomposed into local and nonlocal parts~\cite{Stepanov16-2},
\begin{align}
   \Pi^\alpha_q=&\pi^\alpha_\omega+\sum_{\nu}\Lambda^{i,\alpha}_{\nu q}\tilde{X}^0_{\nu}(q)\bar{\lambda}^{i,\alpha}_{\nu\omega}
    =\pi^\alpha_\omega+\tilde{\Pi}^\alpha_q.\label{eq:pitext}
\end{align}
The nonlocal corrections are denoted as $\tilde{\Pi}$, analogous to the dual boson formula for the susceptibility~\eqref{eq:suscdb} and
$\tilde{X}^0_\nu(q)$ is a nonlocal bubble,
\begin{align}
    \tilde{X}^0_\nu(q)=\sum_\kv\tilde{G}_{k+q}\tilde{G}_k.\label{eq:nonlocalbubble}
\end{align}
Here, $\tilde{G}_k=G_k-g_\nu$ is the nonlocal DMFT Green's function, which decays with the frequency as ${1}/{\nu^2}$,
and $\Lambda^{i,\alpha}$ is the left-sided lattice Hedin vertex. Equation~\eqref{eq:pitext} is depicted diagrammatically in Fig.~\ref{fig:diagrams} b).

We now come to the main result, which is a nonlocal ladder equation for the Hedin vertex in the DMFT approximation~\cite{Katanin09}.
It is shown in Appendix~\ref{app:hedin} that,
\begin{align}
    \Lambda^{i,\alpha}_{\nu q}=&\lambda^{i,\alpha}_{\nu\omega}+\sum_{\nu'}\Lambda^{i,\alpha}_{\nu' q}\tilde{X}^0_{\nu'}(q)f^{i,\alpha}_{\nu'\nu\omega},
    \label{eq:liladder}
\end{align}
which is depicted in Fig.~\ref{fig:diagrams} a).
Note that $f^{i}$ is the $U^\alpha$-\textit{irreducible} four-leg vertex of the impurity model, it is the only true four-point object needed in the calculation.

\subsection{Comparison to dual boson formula}\label{sec:pixcomparison}
It will now be shown that the formula~\eqref{eq:pitext} for the polarization is numerically more efficient than the dual boson formula~\eqref{eq:suscdb}.
To this end, let us recall that in the latter case the nonlocal corrections are given as
[see Appendix~\ref{app:susc}, Refs.~\cite{vanLoon14-2,Hafermann14-2}, and Fig.~\ref{fig:diagrams} c)],
\begin{align}
    \tilde{X}^\alpha_q=2\sum_{\nu}\Lambda^{\alpha}_{\nu q}\tilde{X}^0_{\nu}(q)\bar{\lambda}^{\alpha}_{\nu\omega},
\end{align}
similar to $\tilde{\Pi}$ in Eq.~\eqref{eq:pitext}, except that the $U$-\textit{reducible} three-leg vertices $\Lambda,\lambda$
are in place of the Hedin vertices $\Lambda^i,\lambda^i$ (and the factor $2$).
Furthermore, the vertex $\Lambda$ of the lattice is given by the same ladder equation~\eqref{eq:liladder} [see also Fig.~\ref{fig:diagrams} a)],
albeit the label '$i$' needs to be omitted, and there is hence a complete formal analogy in the calculation of $\Pi$ and $X$.

Let us compare the first four-point vertex contribution to the nonlocal correction terms $\tilde{X}$ and $\tilde{\Pi}$,
by expanding the ladder equations for the three-leg vertices $\Lambda$ and $\Lambda^i$, respectively, see also Eq.~\eqref{eq:liladder},
\begin{align}
    \tilde{X}_q/2\;(\text{or}\;\tilde{\Pi}_q)
    =&\sum_{\nu}\lambda^{(i)}_{\nu\omega}\tilde{X}^0_{\nu}(q)\bar{\lambda}^{(i)}_{\nu\omega}\label{eq:expand}\\
    +&\sum_{\nu\nu'}\lambda^{(i)}_{\nu\omega}\tilde{X}^0_{\nu}(q)f^{(i)}_{\nu\nu'\omega}
    \tilde{X}^0_{\nu'}(q)\bar{\lambda}^{(i)}_{\nu'\omega}+...,\notag
\end{align}
where the flavor label $\alpha$ was omitted for readability.

Typically the calculation of the impurity three-leg vertices $\lambda^{(i)}$ is more efficient than that of the four-leg vertices $f^{(i)}$,
in the latter case one likes to minimize the domain of measurement for $\nu,\nu',\omega$.
The question is therefore how the cutoff error in the four-point corrections that arise in the second line of Eq.~\eqref{eq:expand} affects the calculation.
It is useful to analyze the convergence of the term that is written out in the second line of Eq.~\eqref{eq:expand},
let us consider first the limit $|\nu|\rightarrow\infty$ while $\nu'$ and $\omega$ are kept constant:

According to Eq.~\eqref{eq:lambdairrtext} the decay of the vertices $\lambda_{\nu\omega}$ and $\lambda^i_{\nu\omega}$ with the
frequency $\nu$ is the same except for a prefactor $[1-U^\alpha\pi^\alpha_\omega]^{-1}$,
therefore, the difference in the three-leg vertices does not lead to a different convergence of the $\nu$-summations in $\tilde{X}$ and $\tilde{\Pi}$.
Also in both cases the nonlocal bubble $\tilde{X}^0_{\nu}(q)$ defined in Eq.~\eqref{eq:nonlocalbubble} decays as $1/\nu^{4}$.
However, the vertices $f$ and $f^i$ behave differently, which follows from an observation in Ref.~\cite{Rohringer16}:
In the limit $|\nu|\rightarrow\infty$ all diagrams contributing to $f$ that depend on $\nu$ have decayed,
and hence asymptotically this vertex is given by the diagrams that do \textit{not} depend on $\nu$ at all.
According to the argument in the reference these diagrams are all $U$-reducible, one can write for \textit{fixed} $\nu'$,
\begin{align}
    f^\alpha_{\nu\nu'\omega}=&U^\alpha+U^\alpha\sum_{\nu_1}g_{\nu_1} g_{\nu_1+\omega}f^\alpha_{\nu_1\nu'\omega}+\mathcal{O}\left(\frac{1}{\nu}\right).
\end{align}
Factoring out $U^\alpha$ one identifies the reducible three-leg vertex $\lambda$ [see below Eq.~\eqref{eq:3pvertex}], therefore,
\begin{align}
    \lim\limits_{|\nu|\rightarrow\infty}f^\alpha_{\nu\nu'\omega}=&\,U^\alpha{\lambda}^\alpha_{\nu'\omega}
    =w^{\alpha}_\omega{\lambda}^{i,\alpha}_{\nu'\omega}.\label{eq:fnu}
\end{align}
In the last step Eq.~\eqref{eq:3pvertex} and $w^\alpha=U^\alpha/(1-U^\alpha\pi^\alpha_\omega)$ were used.
Let us now compare to the exact relation between the vertices $f$ and $f^i$ in Eq.~\eqref{eq:firrtext}.
The asymptotic limit of $f$ in Eq.~\eqref{eq:fnu} is given exactly by the asymptotic limit of the $U$-reducible diagrams
$\bar{\lambda}^{i}_{\nu\omega}\,w_\omega\,\lambda^{i}_{\nu'\omega}$ (note that $\bar{\lambda}^i_{\nu\omega}\rightarrow1$ for $|\nu|\rightarrow\infty$).
This is not surprising in view of the observation of Ref.~\cite{Rohringer16} that only $U$-reducible diagrams can be independent of $\nu$.
As a result, the irreducible vertex $f^i_{\nu\nu'\omega}$ decays to zero for $|\nu|\rightarrow\infty$ and \textit{fixed} $\nu'$,
\begin{align}
    f^{i,\alpha}_{\nu\nu'\omega}=&\,0+\mathcal{O}\left(\frac{1}{\nu}\right).
\end{align}
For this reason the four-point corrections in Eq.~\eqref{eq:expand} decay by at least one power of $\nu$ faster
for $\tilde{\Pi}$ than for $\tilde{X}$, which is the central observation of this work.

A comprehensive discussion of the asymptotics of $f$ can be found in Ref.~\cite{Wentzell16}, where it is also shown that in the double limit 
$|\nu|,|\nu'|\rightarrow\infty$ one needs to consider separately the two cases $\nu-\nu'=\text{const}$ and $\omega+\nu+\nu'=\text{const}$,
that is, the elements of $f$ near the main and secondary diagonal.
However, as regards the scope of this work these cases can be ignored, because then the nonlocal bubbles in Eq.~\eqref{eq:expand} decay as
$1/\nu^4$ and $1/(\nu')^4$, respectively, leading to a still faster decay than when only one frequency is large.
In summary, in the dual boson formula each four-point correction comes with a factor $\tilde{X}^0_{\nu}(q)f_{\nu\nu'\omega}$,
which decays like the nonlocal bubble as $1/\nu^4$ due to the constant background of $f$, whereas in the efficient calculation scheme the corrections
enter as $\tilde{X}^0_{\nu}(q)f^{i}_{\nu\nu'\omega}$, which decays at least as $1/\nu^5$ by virtue of the combined decay of nonlocal bubble and vertex $f^i$.

\section{TRILEX-like approximation and two-particle self-consistency}\label{sec:tpscdmf}
This section considers an optimal truncation of the vertex corrections to the Hedin vertex and a two-particle self-consistent constraint on the DMFT susceptibility.

\subsection{TRILEX-like approximation}\label{sec:trilex}
Despite all optimizations it may be unfeasible to take four-point corrections to the Hedin vertex into account,
for example, in multi-orbital settings, cf. Appendix~\ref{app:morb}.
In this case one may consider to neglect vertex corrections in Eq.~\eqref{eq:liladder},
$\Lambda^i\approx\lambda^i$, which is the philosophy of the TRILEX approach.
However, there are two ways to introduce this approximation:
Firstly, in the efficient formula~\eqref{eq:pitext} the local approximation to the Hedin vertex leads to,
\begin{align}
    \Pi^{(2),\alpha}_q=&\pi^\alpha_\omega+\sum_{\nu}\lambda^{i,\alpha}_{\nu\omega}\tilde{X}^0_{\nu}(q)\bar{\lambda}^{i,\alpha}_{\nu\omega},\label{eq:trilex2}
\end{align}
which corresponds to replacing the full triangle in Fig.~\ref{fig:diagrams} b) with a bare triangle and one is left with \textit{two} bare triangles.
Secondly, a more direct way to apply the approximation is to insert it into the relation $\Pi^\alpha_q=\sum_k\Lambda^{i,\alpha}_{\nu q}G_k G_{k+q}$,
which is equivalent to equation~\eqref{eq:pitext} when vertex corrections are kept (cf. Appendix~\ref{app:susc}).
Nevertheless, $\Lambda^i\approx\lambda^i$ leads to a different approximation,
\begin{align}
  \Pi^{(1),\alpha}_q=&\sum_{k}\lambda^{i,\alpha}_{\nu\omega}G_k G_{k+q}\notag\\
    =&\pi^\alpha_\omega+\sum_{\nu}\lambda^{i,\alpha}_{\nu\omega}\tilde{X}^0_{\nu}(q),\label{eq:trilex1}
\end{align}
In the second line the nonlocal bubble~\eqref{eq:nonlocalbubble} was introduced using the relation
$\sum_{\kv}G_k G_{k+q}=\tilde{X}^0_{\nu}(q)+g_\nu g_{\nu+\omega}$ and the exact impurity polarization was identified,
$\pi^\alpha_\omega=\sum_{\nu}\lambda^{i,\alpha}_{\nu\omega}g_\nu g_{\nu+\omega}$.
Equation~\eqref{eq:trilex1} corresponds to the way the local approximation is introduced in the TRILEX approach~\cite{Ayral15-2}, it has only \textit{one} bare triangle.

The obvious question is whether the first or the second option is a more viable way to truncate the vertex corrections.
This question can be decided by considering the strong coupling limit, which shows that only $\Pi^{(2)}$
in equation~\eqref{eq:trilex2} correctly describes the effective exchange:
It is shown in Appendix~\ref{app:strongcoupling} that for very large coupling $U\gg t,T$
the static DMFT spin susceptibility of the half-filled Hubbard model takes the form, see also Ref.~\cite{Otsuki19},
\begin{align}
    X^{\sz}(\qv,\omega=0)\widetilde{=}-\frac{2}{2T-I_\qv}\label{eq:XSC},
\end{align}
where $T$ is the temperature and $I_\qv$ is the effective exchange.
The Appendix shows further that the approximations~\eqref{eq:trilex2} and~\eqref{eq:trilex1} yield different expressions for $I_\qv$,
\begin{align}
    I_\qv^{(2)}=&-\frac{2t^2\gamma_\qv}{(\pi^{\sz}_{\omega=0})^2}\sum_\nu\lambda^{i,\sz}_{\nu,\omega=0}(g_{\nu})^4\bar{\lambda}^{i,\sz}_{\nu,\omega=0},\label{eq:jeff2}\\
    I_\qv^{(1)}=&-\frac{2t^2\gamma_\qv}{(\pi^{\sz}_{\omega=0})^2}\sum_\nu\lambda^{i,\sz}_{\nu,\omega=0}(g_{\nu})^4,\label{eq:jeff1}
\end{align}
respectively, where $t$ is the hopping, $\gamma_\qv$ depends on the dispersion of the lattice, for the square lattice $\gamma_\qv=\cos(q_x)+\cos(q_y)$.
In this case $I_\qv$ is a nearest neighbor interaction, it inherits this property from $\varepsilon_\kv=-2t\gamma_\kv$.
\begin{figure}
  \begin{center}
    \includegraphics[width=0.45\textwidth]{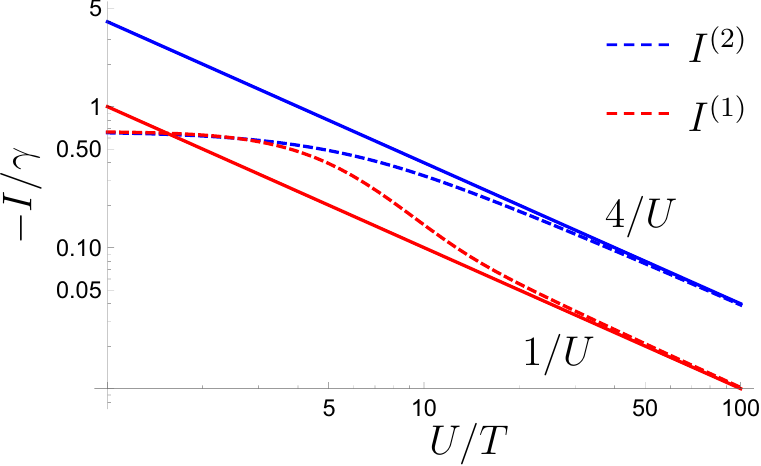}
  \end{center}
    \caption{\label{fig:al} (Color online)  
    Prefactor of the effective exchange in the atomic limit ($\Delta=0$).
    For $U\gg T$ the prefactors corresponding to $I^{(2)}$ and $I^{(1)}$ approach ${4t^2}/{U}$ and ${t^2}/{U}$, respectively, where $t=1$.
    }
\end{figure}

It is instructive to evaluate the impurity quantities that determine $I_\qv$ in the atomic limit where the hybridization function $\Delta$ of DMFT vanishes.
Fig.~\ref{fig:al} shows that for $U\gg T$ one has $-I_\qv^{(2)}/\gamma_\qv\rightarrow4t^2/U$, whereas $-I_\qv^{(1)}/\gamma_\qv\rightarrow t^2/U$.
In combination with equation~\eqref{eq:XSC} this implies that only $I^{(2)}$ recovers the correct N\'eel temperature of
the half-filled Hubbard model on the square lattice $[\qv=(\pi,\pi)]$ in the Heisenberg mean-field limit,
$T_N=\frac{4t^2}{U}$, whereas $I^{(1)}$ is off by a factor $4$.

Apparently, the vertex corrections at each lattice site need to be treated on an equal footing because
the effective exchange is a coupling between equivalent nearest neighbors.
Therefore, approximation~\eqref{eq:trilex2} is used in the applications.
The fact that it recovers the effective exchange implies that four-point vertex corrections
to the efficient formula~\eqref{eq:pitext} can be neglected in the limit $U\gg t,T$, which is confirmed numerically further below.

\subsection{Two-particle self-consistency}
The DMFT susceptibility $X^\sz$ in Eq.~\eqref{eq:susctext} may diverge in two dimensions, in violation of the Mermin-Wagner theorem,
and it shows the mean-field critical behavior near an antiferromagnetic instability in three dimensions~\cite{Rohringer11,Hirschmeier15}.
As discussed in the context of the two-particle self-consistent (TPSC) approach,
these drawbacks are due to the violation of local sum rules~\cite{Vilk97}.
In order to alleviate the mean-field artifacts a frequency-dependent correction is introduced,
\begin{align}
   X^\alpha_q\rightarrow\mathcal{X}^\alpha_q=&\frac{2\Pi^\alpha_{q}}{1-(U^\alpha+\mathcal{U}^{\alpha}_\omega){\Pi}^\alpha_{q}},\label{eq:x2psc}
\end{align}
where $\Pi$ is the DMFT polarization~\eqref{eq:pitext}.
The correction term $\mathcal{U}^\alpha_\omega$ is fixed by the self-consistency condition,
\begin{align}
    \sum_\qv\mathcal{X}^\alpha_q=\chi^\alpha_\omega,\label{eq:tpsc}
\end{align}
thereby $\mathcal{X}$ yields the same kinetic and potential energy as the impurity model of DMFT~\cite{Krien17}.
Furthermore, the local sum rules are satisfied,
\begin{align}
    \sum_q\mathcal{X}^\ch_q=&\sum_\omega\chi^\ch_\omega=-\langle n\rangle-2\langle n_\uparrow n_\downarrow\rangle+\langle n\rangle^2,\label{eq:locch}\\
    \sum_q\mathcal{X}^\sz_q=&\sum_\omega\chi^\sz_\omega=-\langle n\rangle+2\langle n_\uparrow n_\downarrow\rangle,\label{eq:locsp}
\end{align}
which are a manifestation of the Pauli principle ($n^2_\sigma=n_\sigma$, cf. Ref.~\cite{Krien17}).
Note that $\langle n\rangle=2\sum_\nu g_\nu$ and $\langle n_\uparrow n_\downarrow\rangle$ are the density and double occupancy of the impurity model~\eqref{eq:aim}.

The boundedness of $\langle n_\uparrow n_\downarrow\rangle$ in Eq.~\eqref{eq:locsp} prevents the divergence of $\mathcal{X}^\sz_q$ in two dimensions for $T>0$,
because it would lead to the logarithmic divergence of the two-dimensional integral $\sum_\qv$ on the left-hand-side~\footnote{
    This does not directly imply satisfaction of the Mermin-Wagner theorem,
    because it has to be shown in practice that a solution $\mathcal{U}_\omega$ exists that satisfies Eq.~\eqref{eq:tpsc}.}.
For dimensions $d>2$ Eq.~\eqref{eq:locsp} allows magnetic instabilities for $T>0$,
because then the integral $\sum_\qv$ over the divergent integrand remains finite~\cite{Vilk97}.
In the limit $d\rightarrow\infty$ the constraint~\eqref{eq:tpsc} is satisfied by the DMFT susceptibility~\eqref{eq:susctext}
and hence $\mathcal{U}$ is zero in this limit, as expected.
Finally, $\mathcal{X}$ preserves the feature $\imath\omega\mathcal{X}^\alpha_{\qv=\mathbf{0},\omega}=0$
that is satisfied by the conserving DMFT polarization $\Pi$ in the nominator of Eq.~\eqref{eq:x2psc}.
The two-particle spectrum described by $\mathcal{X}$ is therefore ungapped, as required by the global conservation law~\cite{Krien17}~\footnote{
    Despite the ungapped spectrum the Ward identity is nevertheless violated,
    because due to the correction $\mathcal{U}$ the static homogeneous limit of $\mathcal{X}$ is inconsistent
    with the one-particle level of the DMFT approximation~\cite{Krien18}.
}.

For all these reasons the correction $\mathcal{U}_\omega$ in Eq.~\eqref{eq:x2psc} and the constraint~\eqref{eq:tpsc}
appear as suitable in order to remove the mean-field artifacts from the DMFT susceptibility~\eqref{eq:susctext}.
Note that the self-consistency~\eqref{eq:tpsc} does \textit{not} lead to a feedback on the impurity model of DMFT,
which would in general invalidate the conserving features of the polarization~\cite{Krien17}.
The correction $\mathcal{U}_\omega$ is similar to the constant Moriya-$\lambda$ correction~\cite{Katanin09},
it can however not be interpreted straightforwardly as a renormalization of the correlation length, nor is it a retarded interaction.
Instead, one may interpret $\mathcal{U}_\omega$ as an effective vertex correction to the susceptibility,
which takes diagrams beyond DMFT into account that are needed to satisfy the constraint~\eqref{eq:tpsc}.
This interpretation is consistent with the TPSC approach~\cite{Vilk97},
whose non-perturbative features follow due to effective vertex corrections to the RPA susceptibility.
The latter renormalize the mean-field criticality of the RPA~\cite{Dare96}.
Due to the similarities equation~\eqref{eq:x2psc} and the constraint~\eqref{eq:tpsc} are referred to in this work as a two-particle self-consistent dynamical
mean-field (TPSC-DMF) approach to the susceptibility.

\section{Numerical results}\label{sec:results}
The decomposition of the impurity vertex function in Sec.~\ref{sec:impvertices}, the efficient evaluation of the polarization in Sec.~\ref{sec:pilattice},
and the TPSC-DMF approach in Sec.~\ref{sec:tpscdmf} are applied to the two- and three-dimensional Hubbard models~\eqref{eq:hubbard} at half-filling.
In the calculations firstly the DMFT cycle of Sec.~\ref{sec:hubbard} was completed,
then the four- and three-point correlation functions~\eqref{eq:4pvertex} and~\eqref{eq:3pvertex} of the AIM were evaluated,
where a CTQMC solver based on the ALPS libraries~\cite{ALPS2} with improved estimators~\cite{Hafermann12} was used.
The polarization was then evaluated according to Sec.~\ref{sec:pilattice}, then the TPSC-DMF susceptibility was obtained according to Sec.~\ref{sec:tpscdmf}.
The implementation is based on the dual boson code by E.G.C.P van Loon and H. Hafermann~\cite{vanLoon14}.

\subsection{Impurity vertex function}\label{sec:impvertex}
Figure~\ref{fig:pd} further below shows a phase diagram of the three-dimensional Hubbard model.
In this subsection we focus on this model and the values $U/t=6$ and $U/t=14$ of the interaction,
which correspond in DMFT to a bad metal and to an insulator with local moments~\cite{Hirschmeier15}, respectively, the temperature is set to $T/t\approx0.4$.

\begin{figure}[t]
  \begin{center}
    \includegraphics[width=0.49\textwidth]{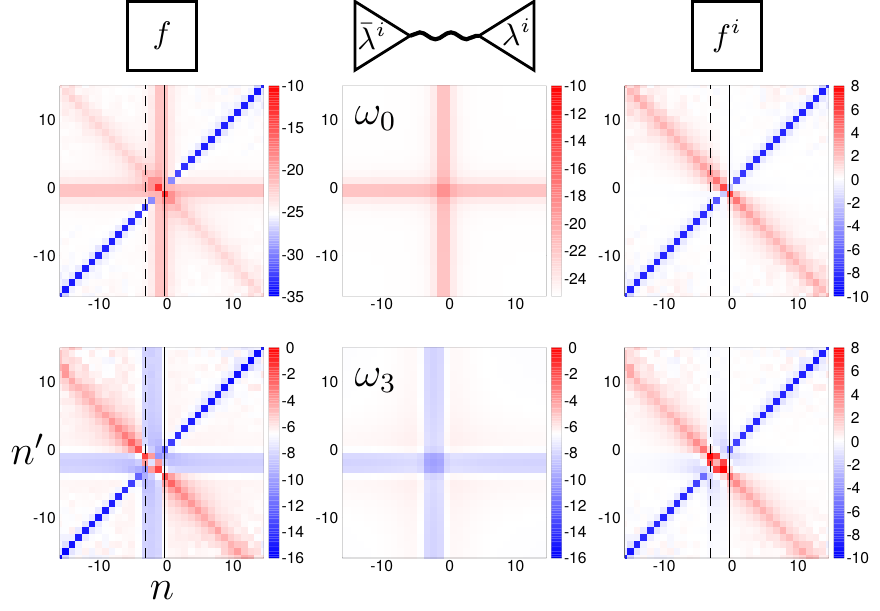}
  \end{center}
    \caption{\label{fig:vertices} (Color online) Real part of the impurity spin vertex $f^{\,\sz}_{\nu\nu'\omega}$ (left) and of its components
    $\bar{\lambda}^{i,\sz}_{\nu\omega}w^\sz_\omega{\lambda}^{i,\sz}_{\nu'\omega}$ (center) and $f^{i,\sz}_{\nu\nu'\omega}$ (right) in a bad metal
    [$U/t=6$, see text] in $\nu_n, \nu'_{n'}$ plane for fixed $\omega_0=0$ (top row) and $\omega_3=6\pi T$ (bottom row).
    White color indicates the constant background, only for $f^{i,\sz}$ on the right this corresponds to zero.
    Notice the smaller Monte Carlo error in the center panels. Vertical lines correspond to cuts ($*$) in Fig.~\ref{fig:cuts}.
    }
\end{figure}

For the metallic regime ($U/t=6$) the left panels of Fig.~\ref{fig:vertices} show the impurity spin vertex function $f^{\,\sz}_{\nu\nu'\omega}$ in the static limit $\omega_0=0$ and for $\omega_3=6\pi T$.
In most directions $f^{\,\sz}$ decays with increasing $\nu,\nu'$ to a constant,
however, it also shows two persistent structures with shapes $+$ and $\times$, see also Ref.~\cite{Rohringer12}.
For finite $\omega_3$ these patterns are shifted along the diagonal.

Important in this work is the exact decomposition $f=f^{i}+\bar{\lambda}^{i}w\lambda^{i}$
discussed in Sec.~\ref{sec:impvertices}. The part that is given by the impurity Hedin vertex $\lambda^{i}$
and by the screened interaction $w$ is shown for $\alpha=\sz$ in the center panels of Fig.~\ref{fig:vertices}.
This object merely shows a $+$ pattern, while the right panels show the $U^\sz$-\textit{irreducible}
vertex $f^{i,\sz}$, which features the $\times$ shape.
This correspondence is also there in the charge channel and in different parameter regimes (not shown).

\begin{figure}[t]
  \begin{center}
    \includegraphics[width=0.49\textwidth]{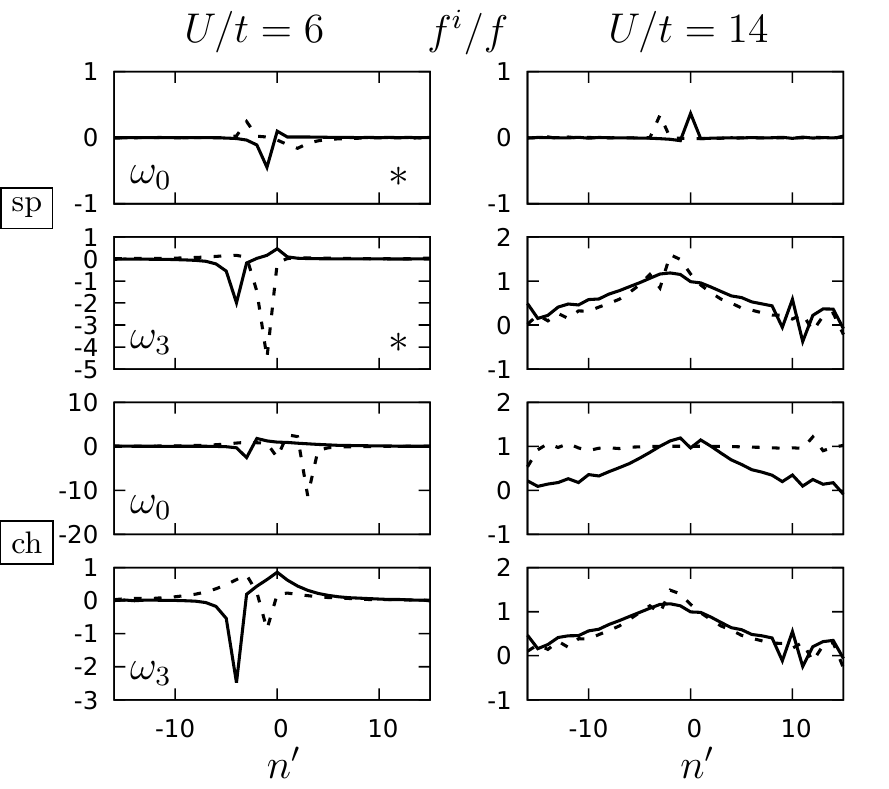}
  \end{center}
    \caption{\label{fig:cuts} (Color online) Ratio $f^{(i)}_{\nu\nu'\omega}/f_{\nu\nu'\omega}$ of irreducible and reducible vertex
    along ${\nu'_{n'}}$ direction, $\nu_n$ and $\omega_m$ are fixed.
    Full and dashed lines show cuts at $\nu_0=\pi T$ and $\nu_{-3}=-5\pi T$, respectively, bosonic frequency ($\omega_0, \omega_3$) as indicated.
    Panels marked with a $*$ correspond to Fig.~\ref{fig:vertices}, where the cuts along $\nu'$-direction are indicated by vertical lines.
    }
\end{figure}

As proven in Sec.~\ref{sec:pixcomparison}, the irreducible vertex $f^i$ does not have a constant background,
and the one of the reducible vertex $f$ originates from the term $\bar{\lambda}^{i}w{\lambda}^{i}$.
The magnitude of this term is determined by the quantity $[1-U^\alpha\pi^\alpha_\omega]^{-1}$, see equation~\eqref{eq:fnu},
which can be very large near a quantum critical point $U^\alpha\pi^\alpha_{\omega=0}\approx1$, where the impurity susceptibility $\chi^\alpha$ is large.
The magnitude of $f$ compared to $f^i$ at large frequencies therefore depends on the physical regime.
For a quantitative comparison Fig.~\ref{fig:cuts} shows the ratio $f^i(\nu,\nu',\omega)/f(\nu,\nu',\omega)$ for fixed $\nu$ and $\omega$ along the $\nu'$-direction.
The left panels of Fig.~\ref{fig:cuts} show the metallic regime $U/t=6$, where the charge and spin susceptibilities $\chi^\ch$ and $\chi^\sz$ are both of non-negligible magnitude,
and hence $f$ is very large compared to $f^i$ at high frequencies.
On the other hand, $\chi^\ch$ is very small in the insulating regime $U/t=14$, and there is no big difference between $f^{\,\ch}$ and $f^{i,\ch}$,
see in particular third panel on the right of Fig.~\ref{fig:cuts}.
Instead, in this regime the static spin susceptibility $\chi^\sz(\omega_0)$ dominates, leading to the fast decay of $f^{i,\sz}/f^{\,\sz}$ visible in the top right panel.
\begin{figure}[b]
  \begin{center}
    \includegraphics[width=0.49\textwidth]{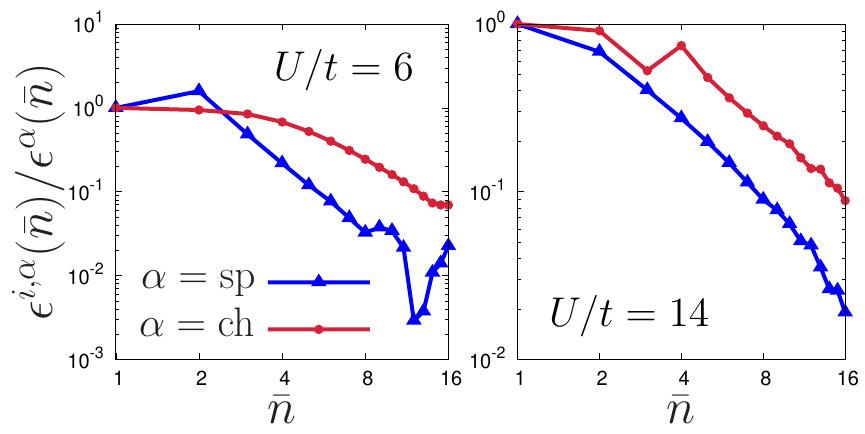}
  \end{center}
    \caption{\label{fig:error} (Color online) Ratio of error functions $\epsilon^i$ and $\epsilon$ as function of cutoff index $\bar{n}$
    for $U/t=6$ (left) and $U/t=14$ (right), corresponds to static vertices in Fig.~\ref{fig:cuts}.
    }
\end{figure}

\subsection{Convergence of frequency summations}
Let us observe the faster convergence of Matsubara summations when the irreducible vertex $f^i$ is used instead of $f$.
For this it is useful to consider the quantity,
\begin{align}
    c^{(i),\alpha}(\bar{n})=\!\!\sum_{n,n'=-\bar{n}}^{\bar{n}-1}\!\!g^{2}(\nu_n)f^{(i),\alpha}(\nu_n,\nu'_{n'},\omega=0)g^{2}(\nu'_{n'}),\notag
\end{align}
which determines the vertex corrections to the static impurity susceptibility $\chi^\alpha_{\omega=0}$ (polarization $\pi^\alpha_{\omega=0}$),
for finite $\bar{n}$ subjected to a cutoff error. A meaningful measure for convergence is
$\epsilon^{(i),\alpha}(\bar{n})=|1-c^{(i),\alpha}(\bar{n}-1)/c^{(i),\alpha}(\bar{n})|$.

Figure~\ref{fig:error} shows the ratio $\epsilon^{i}(\bar{n})/\epsilon(\bar{n})$ as function of the cutoff $\bar{n}$ for the cases discussed in Sec.~\ref{sec:impvertex}.
Clearly, the summation over $f^i$ excels in all cases,
having both the numerically smaller error $\epsilon^i(\bar{n})<\epsilon(\bar{n})$ and the better scaling with $\bar{n}$.
Irregular behavior sets in for large $\bar{n}$ when the Monte Carlo noise exceeds the cutoff error.
Surprisingly, the improvement is even sizable in the charge channel for $U/t=14$, where due to the tiny susceptibility $\chi^\ch(\omega=0)$
the reducible vertex $f^\ch$ has only a small constant background, a worst case scenario.
Nevertheless, for $\bar{n}=16$ (i.e., a $32\times32$ grid) the respective error $\epsilon^{i,\ch}(\bar{n}=16)$
is ten times smaller than $\epsilon^{\ch}(\bar{n}=16)$, see right panel of Fig.~\ref{fig:error}.
In the physically more relevant spin channel this ratio is on the order of one hundred.
One should note that in equation~\eqref{eq:expand} summations converge even faster, thanks to the nonlocal bubble $\tilde{X}^0$.
The improvement of $f^i$ over $f$ in the second line of equation~\eqref{eq:expand} is comparable to or better than the example in Fig.~\ref{fig:error} (not shown).

\begin{figure}
  \begin{center}
    \includegraphics[width=0.49\textwidth]{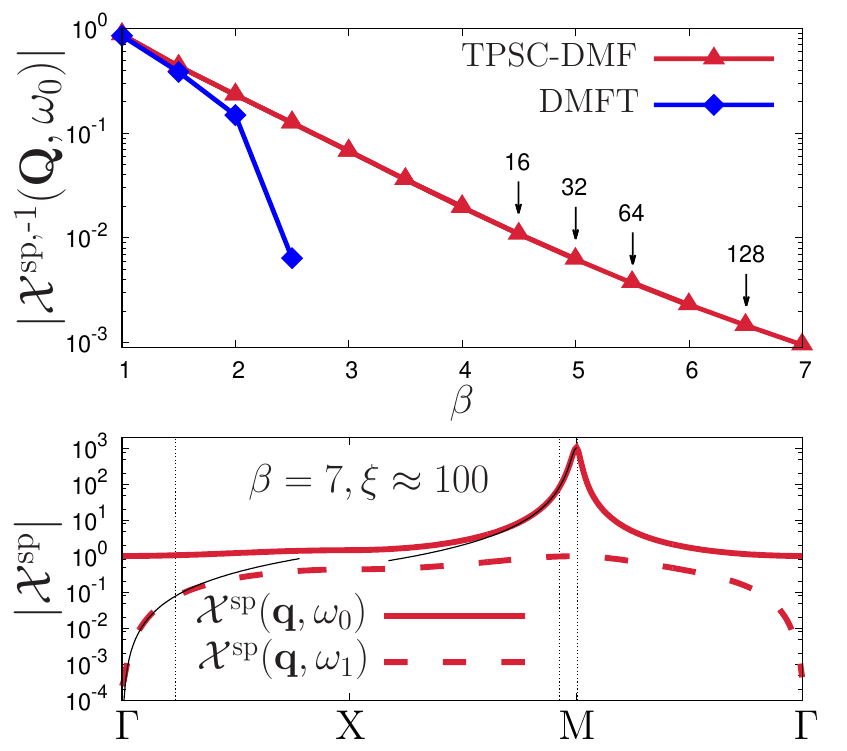}
  \end{center}
    \caption{\label{fig:scaling} (Color online) TPSC-DMF results for the half-filled square lattice at $U/t=8$.
    Top: Scaling of static spin susceptibility at $\Qv=(\pi,\pi)$ with inverse temperature.
    Arrows mark finite-size effects at indicated linear lattice size. 
    Bottom: Static susceptibility (bold red) and at $\omega_1$ (dashed red) in the Brillouin zone at low temperature. 
    Black lines show fits near M and $\Gamma$ [see text], vertical lines indicate fitting intervals.
    }
\end{figure}
\begin{figure}
  \begin{center}
    \includegraphics[width=0.49\textwidth]{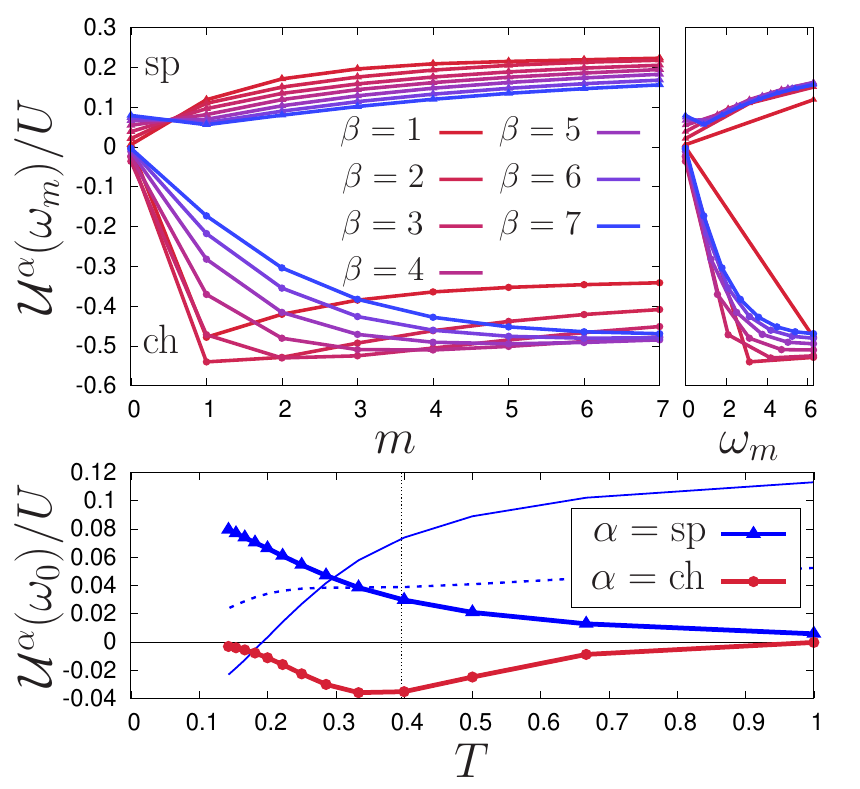}
  \end{center}
    \caption{\label{fig:uomega} (Color online)
    Self-consistent effective vertex correction $\mathcal{U}^\alpha(\omega_m)$ in units of $U$, corresponds to Fig.~\ref{fig:scaling}.
    Top: As function of Matsubara index (left) and frequency (right).
    Bottom: Static components as function of temperature.
    Thin lines show $\mathcal{U}^\sz(\omega_1)-\mathcal{U}^\sz(\omega_0)$ [full] and $\mathcal{U}^\sz(\omega_2)-\mathcal{U}^\sz(\omega_1)$ [dashed],
    vertical line indicates N\'eel temperature of DMFT.
    }
\end{figure}

\subsection{Two-particle self-consistent susceptibility}
The TPSC-DMF susceptibility $\mathcal{X}$ is calculated according to Sec.~\ref{sec:tpscdmf}.
Firstly, it is verified for the Hubbard model on the square lattice that $\mathcal{X}^\sz(\Qv,\omega_0)$ obeys the exponential scaling
with temperature required by the Mermin-Wagner theorem, where $\Qv=(\pi,\pi)$.
This is shown in the top panel of Fig.~\ref{fig:scaling} for $U/t=8$,
at low temperature this corresponds in the DMFT approximation to a strongly correlated Fermi liquid (when paramagnetism is enforced).
With increasing $\beta=\frac{1}{T}$ the DMFT susceptibility $X^\sz$ quickly diverges,
whereas the effective vertex correction $\mathcal{U}$ prevents that the same happens to $\mathcal{X}^\sz=[X^{\sz,-1}-{\mathcal{U}}/{2}]^{-1}$.
For large $\beta$ the correlation length $\xi$ eventually exceeds any fixed system size.
Finite-size effects are noticeable when $\xi$ is of order of the half linear system size,
then the self-consistent calculation of $\mathcal{U}$ becomes inaccurate (arrows).

The bottom panel of Fig.~\ref{fig:scaling} shows $\mathcal{X}^\sz$ in the Brillouin zone
for the largest lattice size $256\times256$ and the lowest considered temperature $T=1/7$.
The figure demonstrates simultaneously features of the Mermin-Wagner theorem and of the conservation law:
On the one hand the static susceptibility $\mathcal{X}^\sz(\qv\approx\Qv,\omega_0)$ shows the required Lorentzian (Ornstein-Zernike) form~\cite{Rohringer16},
while on the other hand $\mathcal{X}^\sz(\qv\approx\mathbf{0},\omega_1)\propto|\qv|^2$, which is required by global spin conservation~\cite{Hafermann14-2}.

The top panels of Fig.~\ref{fig:uomega} show the effective vertex correction $\mathcal{U}^\alpha(\omega_m)$
as function of the Matsubara index $m$ and as function of frequency $\omega_m$.
It is $\mathcal{U}^\sz(\omega_0)>0$, which is required in order for $[-U+\mathcal{U}^\sz(\omega_0)]\Pi^\sz(\Qv,\omega_0)<1$,
preventing the divergence of $\mathcal{X}^\sz$ [cf. Eq.~\eqref{eq:x2psc}, note that $\mathcal{X},\Pi<0$].
The temperature dependence of the static spin component $\mathcal{U}^\sz(\omega_0)$ is drawn in the bottom panel of Fig.~\ref{fig:uomega},
it is consistent with a smooth crossover from a high temperature regime above the N\'eel temperature $T_N\approx0.4$
of DMFT into a low temperature regime, which is located roughly below $T_X\sim0.25$.
Below this temperature the finite size effects documented in the top panel of Fig.~\ref{fig:scaling} indicate a fast increase of the correlation length,
consistent with a renormalized classical regime~\cite{Vilk97}.
A change in the temperature dependence of $\mathcal{U}^\sz(\omega_0)$ here is plausible,
because the momentum integration $\sum_\qv\mathcal{X}(\qv,\omega_0)$
that enters the TPSC-DMF self-consistency~\eqref{eq:tpsc} is increasingly dominated by the Lorentzian centered at the M point,
see bold red line in bottom panel of Fig.~\ref{fig:scaling}, whereas at high temperature also other parts of the Brillouin zone contribute.
The magnitude of $T_X$ corresponds very well to TPSC results at smaller interaction~\cite{Vilk97}.

The corrections $\mathcal{U}^\sz(\omega>0)$ to the dynamical susceptibility are not affected by $T_N$.
Indeed, the dashed red line in the bottom panel of Fig.~\ref{fig:scaling} exemplifies that the dynamical susceptibility $\mathcal{X}(\qv,\omega>0)$
remains flat even far below the N\'eel temperature of DMFT, which is therefore not a special point.
Due to the different temperature dependence of its static and dynamic components $\mathcal{U}^\sz(\omega)$ develops a kink
and $\mathcal{U}^\sz(\omega_1)-\mathcal{U}^\sz(\omega_0)$ changes sign near $T_X$, see bottom panel of Fig.~\ref{fig:uomega}.
In contrast, $\mathcal{U}^\sz(\omega_2)-\mathcal{U}^\sz(\omega_1)$ is largely independent of temperature over a wide range,
although it does show a downturn at very low temperature.
Weak temperature dependence of $\mathcal{U}^\sz(\omega>0)$ was also observed in the three-dimensional case discussed in the following section.

Also the effective vertex correction $\mathcal{U}^\ch(\omega)<0$ of the charge channel is drawn in Fig.~\ref{fig:uomega}.
The bottom panel shows that its static component is significant only in a region around the N\'eel temperature of DMFT.
Interestingly, it seems therefore that static charge correlators of DMFT, such as the compressibility, remain asymptotically unrenormalized at low temperature.
On the other hand, the top left panel of Fig.~\ref{fig:uomega} shows that the dynamic part $\mathcal{U}^\ch(\omega>0)$
is mostly on the order of half the Hubbard interaction $U$, indeed a very large correction.

As function of $\omega$ both $\mathcal{U}^\ch(\omega)$ and $\mathcal{U}^\sz(\omega)$ approach a constant,
reminiscent of the Moriya-$\lambda$ correction and of the self-consistent dual boson approach~\cite{Stepanov16}.
The sign of these corrections is consistently the opposite of $U^\ch=+U$ and $U^\sz=-U$, respectively, which may be interpreted as a screening.
Due to the frequency dependence of $\mathcal{U}(\omega)$ the criticality of static quantities does not affect dynamic ones.
This is different from TPSC and Moriya-$\lambda$, where the same self-consistent correction enters the susceptibility at all frequencies equally.

\subsection{Criticality in three dimensions}
\begin{figure}
  \begin{center}
    \includegraphics[width=0.49\textwidth]{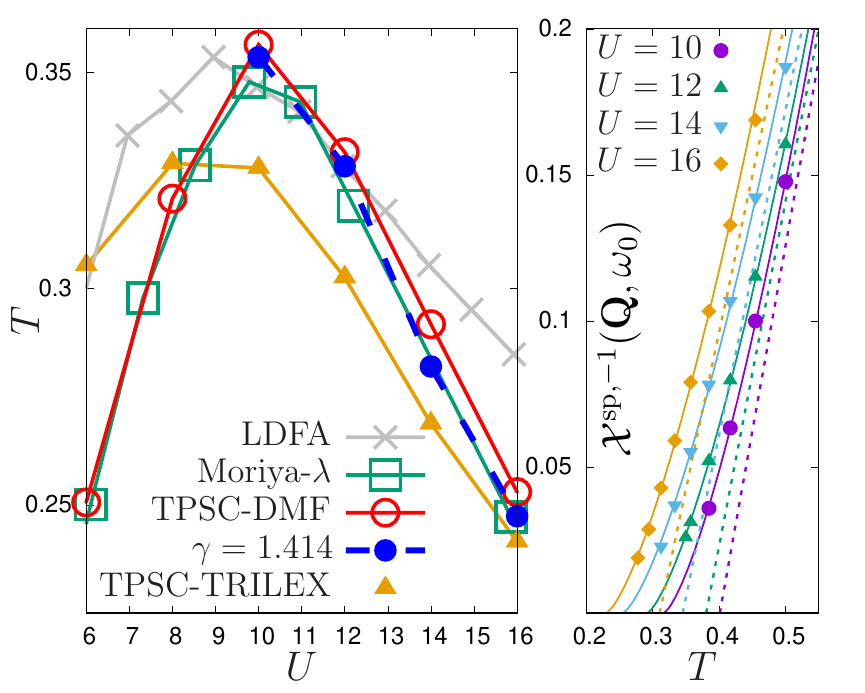}
  \end{center}
    \caption{\label{fig:pd} (Color online) Left: N\'eel temperature of $d=3$ Hubbard model.
    TPSC-DMF results with fixed $\gamma\approx1.4$ (full circles) and with $\gamma$ as free parameter (open circles) are shown.   
    Reprinted gray and green data points with permission from [Rohringer et al., Rev. Mod. Phys, 90, 025003 (2018), see Fig. 22].
    Copyright (2018) by the American Physical Society.
    Right: TPSC-DMF susceptibility \textit{without} vertex corrections $f^i$.
    Dashed lines indicate extrapolation of high-$T$ mean-field behavior, bold lines show fit function $a(T-T_c)^{1.414}$.}
\end{figure}

A further benchmark for the TPSC-DMF susceptibility is to consider criticality when a spontaneous phase transition is indeed allowed,
as is the case in the half-filled three-dimensional Hubbard model.
Figure~\ref{fig:pd} shows the N\'eel temperature predicted by the ladder dual fermion approach (LDFA) 
and by the Moriya-$\lambda$-corrected DMFT susceptibility~\cite{Rohringer17}.
The figure also shows the phase boundary predicted by the TPSC-DMF susceptibility, where $\mathcal{X}^{\sz,-1}(\Qv,\omega_0)$, $\Qv=(\pi,\pi,\pi)$
was fitted with the function $a(T-T_c)^{-\gamma}$ in order to obtain the critical temperature $T_c$ and the critical exponent $\gamma$.
The fit interval needs to be bounded from above by the high-$T$ mean-field regime and from below by finite size effects.
The upper bound was determined as in Ref.~\cite{Rohringer11},
the lower bound is the temperature where the correlation length $\xi$ exceeds $1/6$ of the linear system size of the $16\times16\times16$ lattice, as in Ref.~\cite{Hirschmeier15}.
The boundary obtained by fitting $a,T_c$, and $\gamma$ is in excellent agreement with the Moriya-$\lambda$ correction.

The maximum of $T_c$ at $U/t=10$ marks the crossover from the bad metal to the insulating regime~\cite{Hirschmeier15}.
It was found that already at this point the three-dimensional Hubbard model exhibits the Heisenberg universality class~\cite{Rohringer11}, where $\gamma\approx1.4$.
Consistent with this the fit of $\mathcal{X}$ for $U/t\geq10$ yields an exponent of roughly $1.35$, which compares to the mean-field exponent $1$ of DMFT. 
In this regime $T_c$ was also estimated with $\gamma$ assumed to be known from the Heisenberg model, see blue circles in Fig.~\ref{fig:pd},
which leads to an even better agreement with the Moriya-$\lambda$ correction, it therefore seems that the TPSC-DMF approach predicts the same critical behavior~\footnote{
    The similar results are a consequence of similar self-consistency conditions.
    The Moriya-$\lambda$ is fixed by the local sum rules~\eqref{eq:locch} and~\eqref{eq:locsp},
    whose left-hand-sides are in general dominated by the static term $\omega=0$ near a phase transition,
    in this case $\mathcal{U}^\alpha(\omega=0)\approx\lambda^\alpha_\text{Moriya}$.
}.

Lastly, $T_c$ was also determined when vertex corrections to the Hedin vertex are neglected, $\Lambda^i\approx\lambda^i$.
This approximation is applied as in equation~\eqref{eq:trilex2}, $\Pi\approx\Pi^{(2)}$, for the reasons explained in Sec.~\ref{sec:trilex}.
Note that once again the constraint $\mathcal{X}_\text{loc}=\chi$ is satisfied by self-consistent adjustment of $\mathcal{U}$ in Eq.~\eqref{eq:x2psc}.
In fact, also this approximation clearly deviates from the mean-field criticality near the transition and for $U/t\geq10$
is well-described by the Heisenberg critical exponent, as shown in the right panel of Fig.~\ref{fig:pd}.
Without vertex corrections $T_c$ lies reasonably close to the result \textit{with} the vertex corrections (left panel, yellow and red lines)
but the deviation depends on the physical regime.
For large coupling the vertex corrections have negligible influence on $T_c$,
which confirms the analytical result of Sec.~\ref{sec:trilex}, but they play an important role in the region where DMFT predicts a bad metal.

\section{Conclusions}\label{sec:conclusions}
An efficient method to evaluate the DMFT susceptibility was presented by making use of the Hedin three-leg vertex.
Vertex corrections to the latter arise in the form of a four-point vertex $f^i$ of the Anderson impurity model
that is irreducible with respect to the bare interaction $\pm U$. This vertex has no constant background, in contrast to the full impurity vertex $f$.
Furthermore, the ladder equation for the Hedin vertex is formulated in terms of nonlocal Green's functions, as in the dual fermion approach~\cite{Rubtsov08}.
The combination of the fast decay of the nonlocal Green's functions with the decay of the irreducible vertex $f^i$ leads to
a faster convergence of frequency summations compared to the dual fermion and dual boson approaches~\cite{Rubtsov12}.
As a result, the measurement of the four-point vertex can be restricted to a smaller frequency window.
The efficient calculation scheme can be generalized to multi-orbital Hubbard models and symmetry-broken phases (see Appendix~\ref{app:morb}),
furthermore, it may be possible to incorporate it into the dual fermion and dual boson formalisms~\cite{Rubtsov08,Rubtsov12,Stepanov16-2}.

The efficient calculation scheme implicitly takes vertex asymptotics into account, which were discussed, for example,
in Refs.~\cite{Kunes11,Wentzell16,Kaufmann17,Tagliavini18}. In the implementation it is nevertheless not necessary to consider the large frequency limits explicitly,
because the contributions to the DMFT susceptibility that originate from the constant background of the reducible vertex $f$ are handled in an exact way.
The main difference to the previously presented approaches to reduce the cutoff error by taking vertex asymptotics into account is that
a diagrammatic decomposition of $f$ is employed that is exact for all frequencies, leading to a particularly simple calculation scheme.
The cutoff error may be reduced further by taking the asymptotic behavior of the irreducible vertex $f^i$ into account.

The mean-field instability of the DMFT susceptibility was removed by introduction of a frequency-dependent correction $\mathcal{U}(\omega)$ that is fixed by adjusting the
local susceptibility to the impurity, $\mathcal{X}_\text{loc}(\omega)=\chi(\omega)$.
This approach ensures an ungapped two-particle spectrum and the expected critical behavior in two dimensions in agreement with the Mermin-Wagner theorem,
reminiscent of the two-particle self-consistent (TPSC) approach that is based on the Hartree/RPA approximation~\cite{Vilk97}.
Indeed, the temperature dependence of $\mathcal{U}(\omega=0)$ shows a crossover to a renormalized classical regime, a hallmark effect of the TPSC approach~\cite{Vilk97}.
In the half-filled three-dimensional Hubbard model the criticality of the approach
is consistent with the similar Moriya-$\lambda$ correction used in the dynamical vertex approximation~\cite{Katanin09},
which leads to a renormalized correlation length.

The interpretation of $\mathcal{U}(\omega)$ is however different as a somewhat intransparent vertex correction beyond DMFT,
it is therefore necessary to consider the domain of validity of the approach:
To do this for the weak coupling limit, one may recall that the TPSC approach requires that the Hartree approximation provides a reasonable description of the Fermi surface nesting~\cite{Vilk97}.
However, in the half-filled two-dimensional Hubbard model on the square lattice a pseudogap opens at low temperature due to antiferromagnetic fluctuations~\cite{Vilk96,Schaefer15,vanLoon18-2}.
In this case neither the Hartree approximation nor DMFT provide a good starting point, because they predict a homogeneous Fermi surface with strong nesting.
On the other hand, even when the feedback of the pseudogap on the two-particle spectrum is taken into account it leads to similar results
as the Moriya-$\lambda$-corrected DMFT susceptibility~\cite{Tanaka18,Rohringer17}.
In the large coupling limit the self-consistency $\mathcal{X}_\text{loc}=\chi$ imposes the unscreened local moment of a Mott insulator by construction,
although in reality it may be screened due to short-ranged correlations.
Two-particle self-consistency can therefore impose a bias towards the physics of the impurity model,
furthermore, when it makes a feedback on the impurity model it can violate conservation laws~\cite{Krien17}, which was therefore avoided.
In the future it may be investigated whether the $\mathcal{U}(\omega)$
correction yields a similar feedback on the single-particle spectrum as the Moriya-$\lambda$ correction~\cite{Rohringer16}
and whether it can be generalized to the multi-orbital case~\cite{Galler17}.
A further perspective is to consider the effect of the frequency dependence of $\mathcal{U}(\omega)$ on the two-particle spectrum.

Finally, it was shown that for large coupling vertex corrections to the Hedin vertex play a minor role for the
N\'eel temperature of the half-filled three-dimensional Hubbard model.
This strengthens the case for a local approximation to the Hedin vertex in this regime, as in the TRILEX approach~\cite{Ayral15-2}.
However, it was found that at the level of DMFT the polarization diagram of TRILEX underestimates the prefactor of the effective exchange with energy scale $t^2/U$.
The correct prefactor is obtained when the local approximation to the Hedin vertex is applied to the efficient formula for the polarization,
which corresponds to the dual boson approach~\cite{Stepanov16-2}. This formula treats vertex corrections at each lattice site on an equal footing.

During the completion of this work a manuscript was preprinted~\cite{Otsuki19}
that derives a strong coupling form of the DMFT spin susceptibility with an effective exchange cutoff.
Here this quantity was expressed in terms of local Hedin vertex and polarization of the impurity model.
The latter remain finite at zero temperature, the effective exchange is therefore well-defined in this limit.
The calculation of the spin susceptibility in the Mott phase at zero temperature is an unsolved problem~\cite{Georges96,Guerci18,Krien18}.

\acknowledgments
I thank the anonymous referees for constructive comments that improved this work.
F.K. thanks E.G.C.P. van Loon and A. Valli for their reading of the manuscript, and A.I. Lichtenstein, E.G.C.P. van Loon, and H. Hafermann for long-time support,
and M. Capone, A. Toschi, K. Held, E.A. Stepanov, J. Otsuki, A. Katanin, and L. Fanfarillo for fruitful discussions.
F.K. acknowledges support by MIUR through the PRIN 2015 program (Prot. 2015C5SEJJ001)
and the SISSA/CNR project "Superconductivity, Ferroelectricity and Magnetism in bad metals" (Prot. 232/2015).

\appendix
\section{$U^\alpha$-irreducible vertices}\label{app:irr}
It is shown how diagrams that are reducible with respect to the bare interaction $U^\alpha$
can be separated from the three-leg vertex $\Lambda$ and from the vertex function $F$,
following an approach of Hertz and Edwards~\cite{Hertz73}.
The relations in this section of the appendix are formally exact for the \textit{paramagnetic} Hubbard model~\eqref{eq:hubbard},
for the Anderson impurity model~\eqref{eq:aim} capital letters may be replaced by small letters ($\Lambda\rightarrow\lambda, F\rightarrow f$, and so on) and 
four-momenta are replaced by frequencies [$k=(\kv,\nu)\rightarrow\nu, q=(\qv,\omega)\rightarrow\omega$].
Generalizations to more general lattice and impurity models are briefly discussed in Appendix~\ref{app:morb}.
\subsection{Correlation functions}
The four-point function is defined as,
\begin{align}
    G^{(4),\alpha}_{kk'q}\!&
    =\!-\frac{1}{2}\sum_{\sigma_i}s^\alpha_{\sigma_1'\sigma_1}s^\alpha_{\sigma_2'\sigma_2}\!\av{T_\tau c_{k\sigma_1}c^\dagger_{k+q,\sigma_1'}c_{k'+q,\sigma_2}c^\dagger_{k'\sigma_2'}}\notag,
\end{align}
where definitions are as in the main text.
It is convenient to define the generalized susceptibility,
\begin{align}
    X^\alpha_{kk'q}=G^{(4),\alpha}_{kk'q}+2\beta G_{k}G_{k'}\delta_q\delta_{\alpha,\ch},
\end{align}
the latter can be represented in terms of a ladder equation $\hat{X}=\hat{X}^0+\hat{X}^0\hat{\Gamma}\hat{X}$,
where $\Gamma$ is the two-particle self-energy and all quantities denote matrices in the labels $k,k'$ and $X^0_{kk'}=N\beta G_kG_{k+q}\delta_{kk'}$ is the bubble.
Matrix multiplication implies a factor $(N\beta)^{-1}$, the labels $q,\alpha$ are suppressed.

\subsection{$U^\alpha$-irreducible generalized susceptibility}\label{app:gsuscirr}
\begin{figure}
  \begin{center}
    \includegraphics[width=0.4\textwidth]{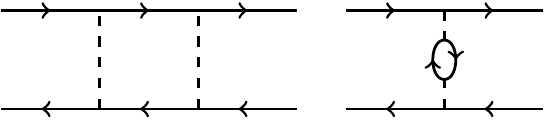}
  \end{center}
    \caption{\label{fig:irr}
    Two diagrammatic contributions to the generalized susceptibility $\hat{X}_q$, dashed lines denote the bare interaction $\pm U$,
    arrows denote Green's function $G$.
    In this work irreducibility implies that removing $U^\alpha$ does not lead to \textit{vertical} separation of a diagram.
    Left: A $U^\alpha$-\textit{reducible} diagram. Right: A $U^\alpha$-\textit{irreducible} diagram.
    }
    \end{figure}

The goal is to separate the diagrams from $\hat{X}$ that are reducible with respect to $U^\ch=+U$ and $U^\sz=-U$, respectively.
To this end, one defines $\hat{\Gamma}^i=\hat{\Gamma}-\hat{\Gamma}^0$, where ${\Gamma}^{0,\alpha}_{kk'}=U^\alpha$ is the \textit{bare} two-particle self-energy.
The ladder equation for $\hat{X}$ can therefore be written as,
\begin{align}
    \hat{X}=&\hat{X}^0+\hat{X}^0(\hat{\Gamma}^i+\hat{\Gamma}^0) \hat{X},\notag\\
    \Leftrightarrow \hat{X}^{0,-1}=&\hat{X}^{-1}+\hat{\Gamma}^i+\hat{\Gamma}^0,\label{eq:gsuscbse}
\end{align}
which implies super-matrix inversion with respect to $k,k'$.
Let us now define the $\Gamma^0$-\textit{irreducible} generalized susceptibility $\hat{\Pi}$,
\begin{align}
    \hat{\Pi}=&\hat{X}^0+\hat{X}^0\hat{\Gamma}^i \hat{\Pi},\notag\\
    \Leftrightarrow \hat{X}^{0,-1}=&\hat{\Pi}^{-1}+\hat{\Gamma}^i.\label{eq:igsuscbse}
\end{align}
There are no diagrams in $\Pi$ that can be separated into two parts by removing a single vertex $\Gamma^0$ [in the sense of Fig.~\ref{fig:irr}].
Subtracting Eq.~\eqref{eq:igsuscbse} from~\eqref{eq:gsuscbse} eliminates $\Gamma^i$ and $X^0$,
\begin{align}
    0=&\hat{X}^{-1}+\hat{\Gamma}^0-\hat{\Pi}^{-1},\notag\\
    \Leftrightarrow \hat{X}=&\hat{\Pi}+\hat{\Pi}\,\hat{\Gamma}^0\hat{X}.
\end{align}
In explicit notation this relation simplifies (the label $\alpha$ remains dropped), 
\begin{align}
    X_{kk'q}=&\Pi_{kk'q}+\sum_{k_1 k_2}\Pi_{kk_1q}\Gamma^{0}{X}_{k_2 k'q}\notag\\
    =&\Pi_{kk'q}+\left(\sum_{k_1}\Pi_{kk_1q}\right)\Gamma^{0}\left(\sum_{k_2}{X}_{k_2 k'q}\right),\label{eq:xfromxi}
\end{align}
where $\Gamma^0=\pm U$, summations imply $(N\beta)^{-1}$.
\subsection{Three-leg vertices and polarization}
$X$ and $\Pi$ will now be related to the left- and right-sided three-leg vertices $\Lambda^{(i)}$ and $\bar{\Lambda}^{(i)}$, using the definitions,
\begin{align}
    \sum_{k}X_{kk'q}=&\Lambda_{k'q}X^0_{k'q},\;\;\;\sum_{k'}X_{kk'q}=X^0_{kq}\bar{\Lambda}_{kq},\label{eq:xfromlambdared}\\
    \sum_{k}\Pi_{kk'q}=&\Lambda^i_{k'q}X^0_{k'q},\;\;\;\sum_{k'}\Pi_{kk'q}=X^0_{kq}\bar{\Lambda}^i_{kq}\label{eq:xfromlambdairr},
\end{align}
where in the second line the $\Gamma^0$-irreducible (Hedin) three-leg vertex $\Lambda^i$ was introduced and $X^0_{kq}=G_{k+q}G_k$ is the bubble.
The reducible and irreducible three-leg vertices are related via Eq.~\eqref{eq:xfromxi}, which is seen by summation over $k'$,
\begin{align}
    \sum_{k'}X_{kk'q}=&\sum_{k'}\Pi_{kk'q}+\sum_{k_1}\Pi_{kk_1q}\Gamma^{0}\sum_{k'k_2}X_{k_2 k'q},\label{app:xsummedright}\\
    \Leftrightarrow X^0_{kq}\bar{\Lambda}_{kq}=&X^0_{kq}\bar{\Lambda}^i_{kq}+X^0_{kq}\bar{\Lambda}^i_{kq}\Gamma^{0}\sum_{k'k_2}X_{k_2 k'q}.
\end{align}
Finally, dividing by $X^0_{kq}$ and identifying the susceptibility, $X_q=2\sum_{kk'}X_{k k'q}$, one arrives at the simple relation,
\begin{align}
    \bar{\Lambda}^\alpha_{kq}=&\bar{\Lambda}^{i,\alpha}_{kq}\left(1+\frac{1}{2}U^\alpha X^\alpha_q\right)\label{app:lambdairrfromx},\\
    =&{\bar{\Lambda}^{i,\alpha}_{kq}}/({1-U^\alpha \Pi^\alpha_q}).\label{eq:lambdairrfrompi}
\end{align}
where the label $\alpha$ was reintroduced. In the second line the polarization was defined,
\begin{align}
    \Pi^\alpha_q={\frac{1}{2}X^\alpha_q}\bigg/\left({1+\frac{1}{2}U^\alpha X^\alpha_q}\right).\label{app:eq:polarization}
\end{align}
By summing Eq.~\eqref{app:xsummedright} over $k$ one sees that,
\begin{align}
    \Pi_q=\sum_{kk'}\Pi_{k k'q}=\sum_kX^0_{kq}\bar{\Lambda}^i_{kq}.
\end{align}
Note that in contrast to the susceptibility $X_q$ a factor $2$ does not occur [see above Eq.~\eqref{app:lambdairrfromx}].
Similar to Eq.~\eqref{eq:lambdairrfrompi} one derives in an analogous way the relation for the \textit{left-sided} three-leg vertex,
\begin{align}
{\Lambda}^\alpha_{kq}={\Lambda}^{i,\alpha}_{kq}/(1-U^\alpha \Pi^\alpha_q)\label{eq:lambdairrfrompileft}.
\end{align}

\subsection{Four-leg vertices and screened interaction}\label{app:fourleg}
Next, also the vertex function $F$ will be expressed in terms of a $\Gamma^0$-irreducible counterpart $F^i$.
To do this, the following relation between the generalized susceptibility $X$ and $F$ will be used,
\begin{align}
    X_{kk'q}=&X^0_{kq}\delta_{kk'}N\beta+X^0_{kq}F_{kk'q}X^0_{k'q},\label{eq:xandfred}\\
    \Pi_{kk'q}=&X^0_{kq}\delta_{kk'}N\beta+X^0_{kq}F^i_{kk'q}X^0_{k'q}.\label{eq:xandfirr}
\end{align}
Inserting these relations into Eq.~\eqref{eq:xfromxi}, and using once again Eqs.~\eqref{eq:xfromlambdared} and~\eqref{eq:xfromlambdairr} leads to,
\begin{align}
    X^0_{kq}F_{kk'q}X^0_{k'q}=X^0_{kq}F^i_{kk'q}X^0_{k'q}+(X^0_{kq}\bar{\Lambda}^i_{kq})\Gamma^0(\Lambda_{k'q}X^0_{k'q}).\notag
\end{align}
Finally, dividing by $X^0_{kq}X^0_{k'q}$ and using Eq.~\eqref{eq:lambdairrfrompileft},
the reducible vertex $F$ can be expressed in terms of the irreducible vertices $F^i$ and $\Lambda^i$,
\begin{align}
    F^\alpha_{kk'q}=F^{i,\alpha}_{kk'q}+\bar{\Lambda}^{i,\alpha}_{kq}W^\alpha_q\Lambda^{i,\alpha}_{k'q},\label{eq:firr}
\end{align}
where the label $\alpha$ was reintroduced and the screened interaction $W$ is defined as,
\begin{align}
    W^\alpha_q={U^\alpha }/({1-U^\alpha \Pi^\alpha_{q}}).\label{eq:wlat}
\end{align}
For the impurity model one makes in Eqs.~\eqref{eq:lambdairrfrompileft},~\eqref{eq:firr}, and~\eqref{eq:wlat} the replacements
$F\rightarrow f$, $\Lambda\rightarrow\lambda$, $W\rightarrow w$, and $\Pi\rightarrow\pi$,
leading to Eqs.~\eqref{eq:lambdairrtext} and~\eqref{eq:firrtext} in the main text.

\section{Ladder equation for the reducible three-leg vertex}\label{app:3leg}
Ladder equations for the reducible and irreducible three-leg vertices $\Lambda$ and $\Lambda^i$ are derived in the DMFT approximation,
where the two-particle self-energy is approximated with the one of the impurity model~\eqref{eq:aim},
$\Gamma^\alpha_{kk'q}=\gamma^\alpha_{\nu\nu'\omega}$~\cite{Georges96,Krien17}.
In this case the Bethe-Salpeter equation for the lattice vertex function $F$ reads,
\begin{align}
    F^\alpha_{\nu\nu'}(q)=\gamma^\alpha_{\nu\nu'\omega}+\sum_{\nu''}\gamma^\alpha_{\nu\nu''\omega}{X}^0_{\nu''}(q)F^\alpha_{\nu''\nu'}(q),\label{eq:bse}
\end{align}
where it was used that for a local two-particle self-energy $\Gamma$ the vertex function does not depend on the momenta $\kv,\kv'$.
$X^0_\nu(q)=\sum_\kv G_{k}G_{k+q}$ denotes the bubble of DMFT Green's functions~\eqref{eq:gf}.

By $\nu,\nu'$-matrix inversion one obtains from Eq.~\eqref{eq:bse} in a short notation,
$\hat{\gamma}^{\alpha,-1}_\omega=\hat{F}^{\alpha,-1}_q+\hat{X}^0(q)$, where $X^0_{\nu\nu'}(q)=\beta X^0_\nu(q)\delta_{\nu\nu'}$.
Similarly, there exists an impurity Bethe-Salpeter equation, $\hat{\gamma}^{\alpha,-1}_\omega=\hat{f}^{\alpha,-1}_\omega+\hat{\chi}^0(\omega)$,
where $f$ denotes the impurity vertex function and $\chi^0_{\nu\nu'}(\omega)=\beta g_\nu g_{\nu+\omega}\delta_{\nu\nu'}$.
Thereby, $\gamma$ is eliminated in favor of $f$, leading to the exact reformulation of Eq.~\eqref{eq:bse},
\begin{align}
    F^\alpha_{\nu\nu'}(q)=f^\alpha_{\nu\nu'\omega}+\sum_{\nu''}f^\alpha_{\nu\nu''\omega}\tilde{X}^0_{\nu''}(q)F^\alpha_{\nu''\nu'}(q),\label{eq:dfbse}
\end{align}
where $\tilde{X}^0_\nu(q)=\sum_\kv (G_kG_{k+q}-g_\nu g_{\nu+\omega})$ is the nonlocal bubble, see also Ref.~\cite{Hafermann14-2}.

In order to arrive at an analogous ladder equation for the three-leg vertex $\Lambda$,
Eq.~\eqref{eq:dfbse} is multiplied by $G_{k'}G_{k'+q}$, summed over $k'$, and $1$ is added on both sides,
\begin{align}
    &1+\sum_{k'}F^\alpha_{\nu\nu'}(q)G_{k'}G_{k'+q}
    =1+\sum_{k'}f^\alpha_{\nu\nu'\omega}G_{k'}G_{k'+q}\notag\\
    +&\sum_{\nu''}f^\alpha_{\nu\nu''\omega}\tilde{X}^0_{\nu''}(q)\sum_{k'}F^\alpha_{\nu''\nu'}(q)G_{k'}G_{k'+q}.
\end{align}
On the left-hand-side (LHS) arises the \textit{right-sided} three-leg vertex, $\bar{\Lambda}_{\nu q}=1+\sum_{k'}F_{\nu\nu'q}G_{k'}G_{k'+q}$,
on the right-hand-side (RHS) $\sum_\kv G_kG_{k+q}=\tilde{X}^0_\nu(q)+g_\nu g_{\nu+\omega}$ is inserted,
\begin{align}
    &\bar{\Lambda}^\alpha_{\nu q}=1+\sum_{\nu'}f^\alpha_{\nu\nu'\omega}g_{\nu'} g_{\nu'+\omega}+\sum_{\nu'}f^\alpha_{\nu\nu'\omega}\tilde{X}^0_{\nu'}(q)\notag\\
    +&\sum_{\nu''}f^\alpha_{\nu\nu''\omega}\tilde{X}^0_{\nu''}(q)\sum_{k'}F^\alpha_{\nu''\nu'}(q)G_{k'}G_{k'+q}.
\end{align}
On the RHS one identifies the \textit{right-sided}
impurity three-leg vertex $\bar{\lambda}_{\nu\omega}=1+\sum_{\nu'}f_{\nu\nu'\omega}g_{\nu'}g_{\nu'+\omega}$,
and $\sum_{\nu''}f^\alpha_{\nu\nu''\omega}\tilde{X}^0_{\nu''}(q)$ is factored out,
\begin{align}
    &\bar{\Lambda}^\alpha_{\nu q}=\bar{\lambda}^\alpha_{\nu\omega}\\
    +&\sum_{\nu''}f^\alpha_{\nu\nu''\omega}\tilde{X}^0_{\nu''}(q)\left(1+\sum_{k'}F^\alpha_{\nu''\nu'}(q)G_{k'}G_{k'+q}\right)\notag.
\end{align}
The term in brackets is again $\bar{\Lambda}$, leading to the ladder equation for the \textit{right-sided} three-leg vertex,
\begin{align}
    &\bar{\Lambda}^\alpha_{\nu q}=\bar{\lambda}^\alpha_{\nu\omega}+\sum_{\nu'}f^\alpha_{\nu\nu'\omega}\tilde{X}^0_{\nu'}(q)\bar{\Lambda}^\alpha_{\nu'q}.
    \label{eq:lambdabseright}
\end{align}
The analogous ladder equation for the \textit{left-sided} three-leg vertex $\Lambda$ follows from the symmetry of the impurity vertex,
$f_{\nu\nu'\omega}=f_{\nu'+\omega,\nu+\omega,-\omega}$,
\begin{align}
    \Lambda^\alpha_{\nu q}=&\lambda^\alpha_{\nu\omega}+\sum_{\nu'}\Lambda^\alpha_{\nu' q}\tilde{X}^0_{\nu'}(q)f^\alpha_{\nu'\nu\omega}.\label{eq:lambdabseleft}
\end{align}

\section{Efficient formulae for susceptibility and polarization}\label{app:susc}
Efficient formulae for the susceptibility and polarization are derived.
The susceptibility may be calculated from the \textit{reducible} three-leg vertex $\Lambda$ as,
\begin{align}
    X^\alpha_q=&-\langle\rho^\alpha_{-q}\rho^\alpha_q\rangle+\beta\langle\rho^\alpha\rangle\langle\rho^\alpha\rangle\delta_q\delta_{\alpha,\ch}\label{app:xdef}\\
    =&2\sum_{k}\Lambda^\alpha_{k q}G_{k}G_{k+q}.\notag
\end{align}
In the DMFT approximation $\Lambda$ does not depend on $\kv$, hence,
$X^\alpha_q=2\sum_{\nu}\Lambda^\alpha_{\nu q}X^0_\nu(q)$, where $X^0_\nu(q)=\sum_\kv G_{k}G_{k+q}$.
This relation will be rewritten as the sum of impurity susceptibility $\chi$ and nonlocal corrections $\tilde{X}$.

To do this, the bubble $X^0$ is expressed in terms of the nonlocal bubble $\tilde{X}^0$ and the impurity bubble $g_\nu g_{\nu+\omega}$,
$X^0_\nu(q)=\tilde{X}^0_\nu(q)+g_\nu g_{\nu+\omega}$,
furthermore, Eq.~\eqref{eq:lambdabseleft} is substituted for the three-leg vertex $\Lambda$,
\begin{align}
     &X^\alpha_q=2\sum_{k}\Lambda^\alpha_{\nu q}G_{k}G_{k+q}\label{eq:suscexp}\\
    =&2\sum_{\nu}\left[\lambda^\alpha_{\nu\omega}+\sum_{\nu'}\Lambda^\alpha_{\nu' q}\tilde{X}^0_{\nu'}(q)f^\alpha_{\nu'\nu\omega}\right]
    \left[\tilde{X}^0_\nu(q)+g_\nu g_{\nu+\omega}\right].\notag
\end{align}
Four terms arise, the impurity susceptibility can be identified, $\chi^\alpha_\omega=2\sum_{\nu}\lambda^\alpha_{\nu\omega}g_\nu g_{\nu+\omega}$.
Furthermore,
\begin{align}
    &2\sum_{\nu}g_\nu g_{\nu+\omega}\sum_{\nu'}\Lambda^\alpha_{\nu' q}\tilde{X}^0_{\nu'}(q)f^\alpha_{\nu'\nu\omega}\label{eq:identify}\\
    =&2\sum_{\nu'}\Lambda^\alpha_{\nu' q}\tilde{X}^0_{\nu'}(q)\bar{\lambda}^\alpha_{\nu'\omega}-2\sum_{\nu'}\Lambda^\alpha_{\nu' q}\tilde{X}^0_{\nu'}(q)\notag, 
\end{align}
where the \textit{right-sided} impurity three-leg vertex was identified,
$\bar{\lambda}^\alpha_{\nu'\omega}=1+\sum_{\nu}f^\alpha_{\nu'\nu\omega}g_\nu g_{\nu+\omega}$
[its trivial part $1$ is canceled by the second term on the RHS of Eq.~\eqref{eq:identify}].
Using these relations in Eq.~\eqref{eq:suscexp} leads to,
\begin{align}
    X^\alpha_q=&\chi^\alpha_\omega+2\sum_{\nu'}\Lambda^\alpha_{\nu' q}\tilde{X}^0_{\nu'}(q)\bar{\lambda}^\alpha_{\nu'\omega}\notag\\
    -&2\sum_{\nu'}\Lambda^\alpha_{\nu' q}\tilde{X}^0_{\nu'}(q)\\
    +&2\sum_{\nu}\lambda^\alpha_{\nu\omega}\tilde{X}^0_\nu(q)+2\sum_{\nu\nu'}\Lambda^\alpha_{\nu' q}\tilde{X}^0_{\nu'}(q)f^\alpha_{\nu'\nu\omega}\tilde{X}^0_\nu(q).\notag
\end{align}
Using the ladder equation~\eqref{eq:lambdabseleft} for $\Lambda$ it is seen that the second line cancels the third, hence,
\begin{align}
    X^\alpha_q=&\chi^\alpha_\omega+2\sum_{\nu'}\Lambda^\alpha_{\nu' q}\tilde{X}^0_{\nu'}(q)\bar{\lambda}^\alpha_{\nu'\omega}
    =\chi^\alpha_\omega+\tilde{X}^\alpha_q,\label{app:eq:susc}
\end{align}
which is the dual boson formula~\eqref{eq:suscdb}~\cite{Rubtsov12,Hafermann14-2}.

A similar relation will be derived for the polarization $\Pi$.
To do this, let us invoke the local analogue of Eq.~\eqref{eq:lambdairrfrompi},
\begin{align}
    \bar{\lambda}^\alpha_{\nu\omega}=&{\bar{\lambda}^{i,\alpha}_{\nu\omega}}/({1-U^\alpha\pi^\alpha_\omega})\label{eq:lambdairrfrompiimp},
\end{align}
where $\bar{\lambda}, \bar{\lambda}^i$, and $\pi$ are the three-leg vertices and the polarization of the impurity.
The latter is related to $\chi$ analogous to Eq.~\eqref{app:eq:polarization},
\begin{align}
    \pi^\alpha_\omega={\frac{1}{2}\chi^\alpha_\omega}\bigg/\left({1+\frac{1}{2}U^\alpha\chi^\alpha_\omega}\right).\label{app:eq:polarizationimp}
\end{align}
Using Eqs.~\eqref{eq:lambdairrfrompiimp},~\eqref{app:eq:polarizationimp} for the impurity quantities,
and Eqs.~\eqref{eq:lambdairrfrompileft},~\eqref{app:eq:polarization} for the lattice quantities in Eq.~\eqref{app:eq:susc} leads to,
\begin{align}
    &\frac{\Pi^\alpha_q}{1-U^\alpha \Pi^\alpha_q}=\frac{\pi^\alpha_\omega}{1-U^\alpha\pi^\alpha_\omega}\\
    +&\frac{1}{1-U^\alpha \Pi^\alpha_q}
    \frac{1}{1-U^\alpha\pi^\alpha_\omega}\sum_{\nu'}\Lambda^{i,\alpha}_{\nu' q}\tilde{X}^0_{\nu'}(q)\bar{\lambda}^{i,\alpha}_{\nu'\omega}.\notag
\end{align}
Multiplication by $1-U^\alpha \Pi^\alpha_q$ and ${1-U^\alpha\pi^\alpha_\omega}$ leads to the desired relation~\eqref{eq:pitext} for the polarization,
\begin{align}
    \Pi^\alpha_q=&\pi^\alpha_\omega+\sum_{\nu'}\Lambda^{i,\alpha}_{\nu' q}\tilde{X}^0_{\nu'}(q)\bar{\lambda}^{i,\alpha}_{\nu'\omega}\label{app:eq:polarizationfromlambda}
\end{align}
Again, compared to Eq.~\eqref{app:eq:susc} a factor $2$ does not occur.

\section{Ladder equation for the Hedin vertex}\label{app:hedin}
Equation~\eqref{eq:lambdabseleft} is now reformulated for the $U^\alpha$-\textit{irreducible} three-leg vertex $\Lambda^i$.
To do this, Eq.~\eqref{eq:lambdairrfrompileft} and its local analogue 
${\lambda}^\alpha_{\nu\omega}={{\lambda}^{i,\alpha}_{\nu\omega}}/({1-U^\alpha\pi^\alpha_\omega})$
are inserted into the ladder equation~\eqref{eq:lambdabseleft} for $\Lambda$,
\begin{align}
    \Lambda^{i,\alpha}_{\nu q}=&\frac{1-U^\alpha \Pi^\alpha_q}{1-U^\alpha\pi^\alpha_\omega}\lambda^{i,\alpha}_{\nu\omega}
    +\sum_{\nu'}\Lambda^{i,\alpha}_{\nu' q}\tilde{X}^0_{\nu'}(q)f^\alpha_{\nu'\nu\omega},\label{eq:lambdabseleftirrinserted}
\end{align}
both sides were multiplied by a factor $1-U^\alpha \Pi^\alpha_q$.
On the RHS appears the reducible impurity vertex function $f$, which will be eliminated in favor of its irreducible counterpart $f^i$
using the local analogue of Eq.~\eqref{eq:firr},
\begin{align}
    f^\alpha_{\nu\nu'\omega}=f^{i,\alpha}_{\nu\nu'\omega}+\bar{\lambda}^{i,\alpha}_{\nu\omega}w^\alpha_\omega\lambda^{i,\alpha}_{\nu'\omega},\label{eq:firrimp}
\end{align}
where $w$ is the screened interaction of the impurity,
\begin{align}
    w^\alpha_\omega={U^\alpha}/({1-U^\alpha\pi^\alpha_{\omega}})\label{eq:wimp}.
\end{align}
Inserting Eq.~\eqref{eq:firrimp} into Eq.~\eqref{eq:lambdabseleftirrinserted} leads to,
\begin{align}
    \Lambda^{i,\alpha}_{\nu q}=&\frac{1-U^\alpha \Pi^\alpha_q}{1-U^\alpha\pi^\alpha_\omega}\lambda^{i,\alpha}_{\nu\omega}
    +\sum_{\nu'}\Lambda^{i,\alpha}_{\nu' q}\tilde{X}^0_{\nu'}(q)f^{i,\alpha}_{\nu'\nu\omega}\notag\\
    +&\sum_{\nu'}\Lambda^{i,\alpha}_{\nu' q}\tilde{X}^0_{\nu'}(q)\bar{\lambda}^{i,\alpha}_{\nu'\omega}w^\alpha_\omega\lambda^{i,\alpha}_{\nu\omega}.
    \label{eq:lambdabseleftirrinserted2}
\end{align}
Using Eqs.~\eqref{eq:wimp} and~\eqref{eq:wlat} the fraction on the RHS can be expressed as $\frac{w^\alpha_\omega}{W^\alpha_q}$.
Furthermore, Eq.~\eqref{app:eq:polarizationfromlambda} can be used to identify in the second line,
$\sum_{\nu'}\Lambda^{i,\alpha}_{\nu' q}\tilde{X}^0_{\nu'}(q)\bar{\lambda}^{i,\alpha}_{\nu'\omega}=\Pi^\alpha_q-\pi^\alpha_\omega$.
Eq.~\eqref{eq:lambdabseleftirrinserted2} thus becomes,
\begin{align}
    \Lambda^{i,\alpha}_{\nu q}=&\frac{w^\alpha_\omega}{W^\alpha_q}\lambda^{i,\alpha}_{\nu\omega}
    +\sum_{\nu'}\Lambda^{i,\alpha}_{\nu' q}\tilde{X}^0_{\nu'}(q)f^{i,\alpha}_{\nu'\nu\omega}\notag\\
    +&(\Pi^\alpha_q-\pi^\alpha_\omega)w^\alpha_\omega\lambda^{i,\alpha}_{\nu\omega}.
    \label{eq:lambdabseleftirrinserted3}
\end{align}
Using the relation~\eqref{eq:wimp} between $w$ and $\pi$, and the relation~\eqref{eq:wlat} between $W$ and $\Pi$ leads to the desired
ladder equation~\eqref{eq:liladder} for the Hedin vertex,
\begin{align}
    \Lambda^{i,\alpha}_{\nu q}=&\lambda^{i,\alpha}_{\nu\omega}+\sum_{\nu'}\Lambda^{i,\alpha}_{\nu' q}\tilde{X}^0_{\nu'}(q)f^{i,\alpha}_{\nu'\nu\omega}.
    \label{eq:lambdairrladder}
\end{align}

\section{General bare interaction}\label{app:morb}
In the Hedin formalism the bosons arise because Green's function lines are contracted at a bare interaction vertex
that does not depend on fermionic momentum-energies $k=(\kv,\nu)$, see Sec.~\ref{sec:hedin}.
This requirement allows for much more general interaction Hamiltonians than considered here.

In particular, the Appendices~\ref{app:irr}-\ref{app:hedin} (i.e., the efficient calculation of the DMFT polarization)
can be generalized to multi-orbital systems and/or symmetry-broken phases.
In these cases a matrix-valued bare interaction of the form $U^{ab}$ enters the Bethe-Salpeter equation,
where $a=(m_1m_2\sigma_1\sigma_2)$ is a superindex of two orbital and two spin indices~\footnote{
   For the single-band Hubbard model $U^{\up\up\dn\dn}=U^{\dn\dn\up\up}=-U^{\dn\up\up\dn}=-U^{\up\dn\dn\up}=U$, the other elements are zero.
}, see also Ref.~\cite{Galler17}.
As in Appendix~\ref{app:gsuscirr} one removes the bare part from the two-particle self-energy,
$\hat{\Gamma}^{i,ab}=\hat{\Gamma}^{ab}-\hat{\Gamma}^{0,ab}$, where $\Gamma^{0,ab}_{kk'}=U^{ab}$. 
One then derives the crucial equation~\eqref{eq:xfromxi}, which becomes a matrix relation with respect to the superindices,
it serves as the vantage point for the remaining calculations.

On an equal footing it seems possible to introduce a TPSC-DMF prescription~\eqref{eq:tpsc}, $\sum_\qv\mathcal{X}^{ab}_q=\chi^{ab}_\omega$,
which is fixed by an effective vertex correction $\mathcal{U}^{ab}(\omega)$~\footnote{
  It is unclear whether a generalization of TPSC-DMF to symmetry-broken phases inherits
  thermodynamic consistency at second order critical points from the DMFT approximation~\cite{Krienthesis}, which may be clarified in future work.
}.

Finally, it is possible to generalize Appendix~\ref{app:irr} to a nonlocal and/or retarded interaction.
However, only the RPA-like vertex $U(\qv,\omega)$ can be separated from the Bethe-Salpeter equation,
not the Fock exchange $U(\kv'-\kv,\nu'-\nu)$, since it depends on the fermionic variables.
Appendices~\ref{app:3leg}-\ref{app:hedin} rely on the DMFT approximation where interaction of lattice and impurity need to be equivalent.

\section{Strong coupling limit}\label{app:strongcoupling}
This appendix considers phase transitions of the half-filled Hubbard model in the strong coupling limit $U\gg T,t$.
Static impurity quantities carry a label `$0$', e.g., $\pi(\omega=0)=\pi_0$, furthermore $q_0=(\qv,\omega=0)$.
\subsection{DMFT}
Near an instability of the static DMFT spin susceptibility $X^\sz(q_0)=2/[\Pi^{\sz,-1}(q_0)-U^\sz]$
one has for the polarization $\Pi^\sz(q_0)=\pi^\sz_0+\tilde{\Pi}^\sz(q_0)\approx\frac{1}{U^\sz}$.
On the other hand, for strong coupling and at half-filling DMFT predicts a Mott insulator with $\chi^\sz_0\propto-\beta$.
Using $\pi^\sz_0=\frac{1}{2}\chi^\sz_0/[1+U^\sz\frac{1}{2}\chi^\sz_0]$ and $U\beta\gg1$
it follows that $\pi^\sz_0\approx\frac{1}{U^\sz}$. Hence, $|\pi^\sz_0|\gg|\tilde{\Pi}^\sz(q_0)|$ and one can expand,
\begin{align}
    X^\sz(q_0)\approx&\frac{2}{\pi^{\sz,-1}_0-\pi^{\sz,-2}_0\tilde{\Pi}^\sz(q_0)-U^\sz}\\
    =&-\frac{2}{-2\chi^{\sz,-1}_0+\pi^{\sz,-2}_0\tilde{\Pi}^\sz(q_0)}.\label{app:xeff}
\end{align}
In the second line $\chi^\sz_0=2/(\pi^{\sz,-1}_0-U^\sz)$ was used.
Defining the effective exchange as $I_\qv=-\pi^{\sz,-2}_0\tilde{\Pi}^\sz(q_0)$ one arrives at the strong coupling form of the DMFT spin susceptibility~\cite{Otsuki19}.
For very large interaction the local moment is fully developed and $\chi^{\sz,-1}_0\approx-T$, leading to equation~\eqref{eq:XSC} in the main text.

\subsection{TRILEX-like approximation}
Let us consider the approximations~\eqref{eq:trilex2} and~\eqref{eq:trilex1} for the polarization $\Pi$.
Both expressions contain the nonlocal bubble $\tilde{X}^0_\nu(q)=\sum_\kv G_{k}G_{k+q}-g_\nu g_{\nu+\omega}$,
which can be simplified in the strong coupling limit using similar steps as in Ref.~\cite{Otsuki19}.
For small hybridization $\Delta\approx0$ one can expand Green's function $G_k\approx g_\nu+g_\nu\varepsilon_\kv g_\nu$, hence,
\begin{align}
    \tilde{X}^0_\nu(q)\approx&\sum_\kv\left(g_\nu g^2_{\nu+\omega}\varepsilon_{\kv+\qv}+g^2_\nu g_{\nu+\omega}\varepsilon_{\kv}+g^2_\nu g^2_{\nu+\omega}\varepsilon_\kv\varepsilon_{\kv+\qv}\right)\notag\\
    =&g^2_\nu g^2_{\nu+\omega}\sum_{\kv}\varepsilon_\kv\varepsilon_{\kv+\qv},\label{app:nonlocalexpand}
\end{align}
where it was used that $\sum_\kv\varepsilon_\kv=\sum_\kv\varepsilon_{\kv+\qv}=0$.
For the dispersion $\varepsilon_\kv=-2t\gamma_\kv$ of the $d$-dimensional hypercubic lattice with $\gamma_\kv=\sum_{i=1}^d\cos(k_i)$ one has $\sum_\kv\varepsilon_\kv\varepsilon_{\kv+\qv}=2t^2\gamma_\qv$.
Using this and Eq.~\eqref{app:nonlocalexpand} yields for the nonlocal part of~\eqref{eq:trilex2} and~\eqref{eq:trilex1}, respectively,
\begin{align}
    \tilde{\Pi}^{(2),\alpha}_q=&2t^2\gamma_\qv\sum_\nu\lambda^{i,\alpha}_{\nu\omega}g^2_\nu g^2_{\nu+\omega}\bar{\lambda}^{i,\alpha}_{\nu\omega},\\
    \tilde{\Pi}^{(1),\alpha}_q=&2t^2\gamma_\qv\sum_\nu\lambda^{i,\alpha}_{\nu\omega}g^2_\nu g^2_{\nu+\omega}.
\end{align}
Inserting into Eq.~\eqref{app:xeff} leads to the expressions $I^{(2)}$ and $I^{(1)}$ for the effective exchange in equations~\eqref{eq:jeff2} and~\eqref{eq:jeff1}.

 \bibliography{main}

\begin{thebibliography}{53}%
\makeatletter
\providecommand \@ifxundefined [1]{%
 \@ifx{#1\undefined}
}%
\providecommand \@ifnum [1]{%
 \ifnum #1\expandafter \@firstoftwo
 \else \expandafter \@secondoftwo
 \fi
}%
\providecommand \@ifx [1]{%
 \ifx #1\expandafter \@firstoftwo
 \else \expandafter \@secondoftwo
 \fi
}%
\providecommand \natexlab [1]{#1}%
\providecommand \enquote  [1]{``#1''}%
\providecommand \bibnamefont  [1]{#1}%
\providecommand \bibfnamefont [1]{#1}%
\providecommand \citenamefont [1]{#1}%
\providecommand \href@noop [0]{\@secondoftwo}%
\providecommand \href [0]{\begingroup \@sanitize@url \@href}%
\providecommand \@href[1]{\@@startlink{#1}\@@href}%
\providecommand \@@href[1]{\endgroup#1\@@endlink}%
\providecommand \@sanitize@url [0]{\catcode `\\12\catcode `\$12\catcode
  `\&12\catcode `\#12\catcode `\^12\catcode `\_12\catcode `\%12\relax}%
\providecommand \@@startlink[1]{}%
\providecommand \@@endlink[0]{}%
\providecommand \url  [0]{\begingroup\@sanitize@url \@url }%
\providecommand \@url [1]{\endgroup\@href {#1}{\urlprefix }}%
\providecommand \urlprefix  [0]{URL }%
\providecommand \Eprint [0]{\href }%
\providecommand \doibase [0]{http://dx.doi.org/}%
\providecommand \selectlanguage [0]{\@gobble}%
\providecommand \bibinfo  [0]{\@secondoftwo}%
\providecommand \bibfield  [0]{\@secondoftwo}%
\providecommand \translation [1]{[#1]}%
\providecommand \BibitemOpen [0]{}%
\providecommand \bibitemStop [0]{}%
\providecommand \bibitemNoStop [0]{.\EOS\space}%
\providecommand \EOS [0]{\spacefactor3000\relax}%
\providecommand \BibitemShut  [1]{\csname bibitem#1\endcsname}%
\let\auto@bib@innerbib\@empty
\bibitem [{\citenamefont {Georges}\ \emph {et~al.}(1996)\citenamefont
  {Georges}, \citenamefont {Kotliar}, \citenamefont {Krauth},\ and\
  \citenamefont {Rozenberg}}]{Georges96}%
  \BibitemOpen
  \bibfield  {author} {\bibinfo {author} {\bibfnamefont {A.}~\bibnamefont
  {Georges}}, \bibinfo {author} {\bibfnamefont {G.}~\bibnamefont {Kotliar}},
  \bibinfo {author} {\bibfnamefont {W.}~\bibnamefont {Krauth}}, \ and\ \bibinfo
  {author} {\bibfnamefont {M.~J.}\ \bibnamefont {Rozenberg}},\ }\href {\doibase
  10.1103/RevModPhys.68.13} {\bibfield  {journal} {\bibinfo  {journal} {Rev.
  Mod. Phys.}\ }\textbf {\bibinfo {volume} {68}},\ \bibinfo {pages} {13}
  (\bibinfo {year} {1996})}\BibitemShut {NoStop}%
\bibitem [{\citenamefont {Gull}\ \emph {et~al.}(2011)\citenamefont {Gull},
  \citenamefont {Millis}, \citenamefont {Lichtenstein}, \citenamefont
  {Rubtsov}, \citenamefont {Troyer},\ and\ \citenamefont {Werner}}]{Gull11}%
  \BibitemOpen
  \bibfield  {author} {\bibinfo {author} {\bibfnamefont {E.}~\bibnamefont
  {Gull}}, \bibinfo {author} {\bibfnamefont {A.~J.}\ \bibnamefont {Millis}},
  \bibinfo {author} {\bibfnamefont {A.~I.}\ \bibnamefont {Lichtenstein}},
  \bibinfo {author} {\bibfnamefont {A.~N.}\ \bibnamefont {Rubtsov}}, \bibinfo
  {author} {\bibfnamefont {M.}~\bibnamefont {Troyer}}, \ and\ \bibinfo {author}
  {\bibfnamefont {P.}~\bibnamefont {Werner}},\ }\href {\doibase
  10.1103/RevModPhys.83.349} {\bibfield  {journal} {\bibinfo  {journal} {Rev.
  Mod. Phys.}\ }\textbf {\bibinfo {volume} {83}},\ \bibinfo {pages} {349}
  (\bibinfo {year} {2011})}\BibitemShut {NoStop}%
\bibitem [{\citenamefont {Rohringer}\ \emph {et~al.}(2012)\citenamefont
  {Rohringer}, \citenamefont {Valli},\ and\ \citenamefont
  {Toschi}}]{Rohringer12}%
  \BibitemOpen
  \bibfield  {author} {\bibinfo {author} {\bibfnamefont {G.}~\bibnamefont
  {Rohringer}}, \bibinfo {author} {\bibfnamefont {A.}~\bibnamefont {Valli}}, \
  and\ \bibinfo {author} {\bibfnamefont {A.}~\bibnamefont {Toschi}},\ }\href
  {\doibase 10.1103/PhysRevB.86.125114} {\bibfield  {journal} {\bibinfo
  {journal} {Phys. Rev. B}\ }\textbf {\bibinfo {volume} {86}},\ \bibinfo
  {pages} {125114} (\bibinfo {year} {2012})}\BibitemShut {NoStop}%
\bibitem [{\citenamefont {Rohringer}\ \emph {et~al.}(2018)\citenamefont
  {Rohringer}, \citenamefont {Hafermann}, \citenamefont {Toschi}, \citenamefont
  {Katanin}, \citenamefont {Antipov}, \citenamefont {Katsnelson}, \citenamefont
  {Lichtenstein}, \citenamefont {Rubtsov},\ and\ \citenamefont
  {Held}}]{Rohringer17}%
  \BibitemOpen
  \bibfield  {author} {\bibinfo {author} {\bibfnamefont {G.}~\bibnamefont
  {Rohringer}}, \bibinfo {author} {\bibfnamefont {H.}~\bibnamefont
  {Hafermann}}, \bibinfo {author} {\bibfnamefont {A.}~\bibnamefont {Toschi}},
  \bibinfo {author} {\bibfnamefont {A.~A.}\ \bibnamefont {Katanin}}, \bibinfo
  {author} {\bibfnamefont {A.~E.}\ \bibnamefont {Antipov}}, \bibinfo {author}
  {\bibfnamefont {M.~I.}\ \bibnamefont {Katsnelson}}, \bibinfo {author}
  {\bibfnamefont {A.~I.}\ \bibnamefont {Lichtenstein}}, \bibinfo {author}
  {\bibfnamefont {A.~N.}\ \bibnamefont {Rubtsov}}, \ and\ \bibinfo {author}
  {\bibfnamefont {K.}~\bibnamefont {Held}},\ }\href {\doibase
  10.1103/RevModPhys.90.025003} {\bibfield  {journal} {\bibinfo  {journal}
  {Rev. Mod. Phys.}\ }\textbf {\bibinfo {volume} {90}},\ \bibinfo {pages}
  {025003} (\bibinfo {year} {2018})}\BibitemShut {NoStop}%
\bibitem [{\citenamefont {van Loon}\ \emph
  {et~al.}(2014{\natexlab{a}})\citenamefont {van Loon}, \citenamefont
  {Hafermann}, \citenamefont {Lichtenstein}, \citenamefont {Rubtsov},\ and\
  \citenamefont {Katsnelson}}]{vanLoon14-2}%
  \BibitemOpen
  \bibfield  {author} {\bibinfo {author} {\bibfnamefont {E.~G. C.~P.}\
  \bibnamefont {van Loon}}, \bibinfo {author} {\bibfnamefont {H.}~\bibnamefont
  {Hafermann}}, \bibinfo {author} {\bibfnamefont {A.~I.}\ \bibnamefont
  {Lichtenstein}}, \bibinfo {author} {\bibfnamefont {A.~N.}\ \bibnamefont
  {Rubtsov}}, \ and\ \bibinfo {author} {\bibfnamefont {M.~I.}\ \bibnamefont
  {Katsnelson}},\ }\href {\doibase 10.1103/PhysRevLett.113.246407} {\bibfield
  {journal} {\bibinfo  {journal} {Phys. Rev. Lett.}\ }\textbf {\bibinfo
  {volume} {113}},\ \bibinfo {pages} {246407} (\bibinfo {year}
  {2014}{\natexlab{a}})}\BibitemShut {NoStop}%
\bibitem [{\citenamefont {Boehnke}\ and\ \citenamefont
  {Lechermann}(2012)}]{Boehnke12}%
  \BibitemOpen
  \bibfield  {author} {\bibinfo {author} {\bibfnamefont {L.}~\bibnamefont
  {Boehnke}}\ and\ \bibinfo {author} {\bibfnamefont {F.}~\bibnamefont
  {Lechermann}},\ }\href {\doibase 10.1103/PhysRevB.85.115128} {\bibfield
  {journal} {\bibinfo  {journal} {Phys. Rev. B}\ }\textbf {\bibinfo {volume}
  {85}},\ \bibinfo {pages} {115128} (\bibinfo {year} {2012})}\BibitemShut
  {NoStop}%
\bibitem [{\citenamefont {Geffroy}\ \emph {et~al.}(2019)\citenamefont
  {Geffroy}, \citenamefont {Kaufmann}, \citenamefont {Hariki}, \citenamefont
  {Gunacker}, \citenamefont {Hausoel},\ and\ \citenamefont
  {Kune\ifmmode~\check{s}\else \v{s}\fi{}}}]{Geffroy18}%
  \BibitemOpen
  \bibfield  {author} {\bibinfo {author} {\bibfnamefont {D.}~\bibnamefont
  {Geffroy}}, \bibinfo {author} {\bibfnamefont {J.}~\bibnamefont {Kaufmann}},
  \bibinfo {author} {\bibfnamefont {A.}~\bibnamefont {Hariki}}, \bibinfo
  {author} {\bibfnamefont {P.}~\bibnamefont {Gunacker}}, \bibinfo {author}
  {\bibfnamefont {A.}~\bibnamefont {Hausoel}}, \ and\ \bibinfo {author}
  {\bibfnamefont {J.}~\bibnamefont {Kune\ifmmode~\check{s}\else \v{s}\fi{}}},\
  }\href {\doibase 10.1103/PhysRevLett.122.127601} {\bibfield  {journal}
  {\bibinfo  {journal} {Phys. Rev. Lett.}\ }\textbf {\bibinfo {volume} {122}},\
  \bibinfo {pages} {127601} (\bibinfo {year} {2019})}\BibitemShut {NoStop}%
\bibitem [{\citenamefont {Toschi}\ \emph {et~al.}(2007)\citenamefont {Toschi},
  \citenamefont {Katanin},\ and\ \citenamefont {Held}}]{Toschi07}%
  \BibitemOpen
  \bibfield  {author} {\bibinfo {author} {\bibfnamefont {A.}~\bibnamefont
  {Toschi}}, \bibinfo {author} {\bibfnamefont {A.~A.}\ \bibnamefont {Katanin}},
  \ and\ \bibinfo {author} {\bibfnamefont {K.}~\bibnamefont {Held}},\ }\href
  {\doibase 10.1103/PhysRevB.75.045118} {\bibfield  {journal} {\bibinfo
  {journal} {Phys. Rev. B}\ }\textbf {\bibinfo {volume} {75}},\ \bibinfo
  {pages} {045118} (\bibinfo {year} {2007})}\BibitemShut {NoStop}%
\bibitem [{\citenamefont {Galler}\ \emph {et~al.}(2017)\citenamefont {Galler},
  \citenamefont {Thunstr\"om}, \citenamefont {Gunacker}, \citenamefont
  {Tomczak},\ and\ \citenamefont {Held}}]{Galler17}%
  \BibitemOpen
  \bibfield  {author} {\bibinfo {author} {\bibfnamefont {A.}~\bibnamefont
  {Galler}}, \bibinfo {author} {\bibfnamefont {P.}~\bibnamefont {Thunstr\"om}},
  \bibinfo {author} {\bibfnamefont {P.}~\bibnamefont {Gunacker}}, \bibinfo
  {author} {\bibfnamefont {J.~M.}\ \bibnamefont {Tomczak}}, \ and\ \bibinfo
  {author} {\bibfnamefont {K.}~\bibnamefont {Held}},\ }\href {\doibase
  10.1103/PhysRevB.95.115107} {\bibfield  {journal} {\bibinfo  {journal} {Phys.
  Rev. B}\ }\textbf {\bibinfo {volume} {95}},\ \bibinfo {pages} {115107}
  (\bibinfo {year} {2017})}\BibitemShut {NoStop}%
\bibitem [{\citenamefont {Hafermann}\ \emph {et~al.}(2012)\citenamefont
  {Hafermann}, \citenamefont {Patton},\ and\ \citenamefont
  {Werner}}]{Hafermann12}%
  \BibitemOpen
  \bibfield  {author} {\bibinfo {author} {\bibfnamefont {H.}~\bibnamefont
  {Hafermann}}, \bibinfo {author} {\bibfnamefont {K.~R.}\ \bibnamefont
  {Patton}}, \ and\ \bibinfo {author} {\bibfnamefont {P.}~\bibnamefont
  {Werner}},\ }\href {\doibase 10.1103/PhysRevB.85.205106} {\bibfield
  {journal} {\bibinfo  {journal} {Phys. Rev. B}\ }\textbf {\bibinfo {volume}
  {85}},\ \bibinfo {pages} {205106} (\bibinfo {year} {2012})}\BibitemShut
  {NoStop}%
\bibitem [{\citenamefont {Gunacker}\ \emph {et~al.}(2016)\citenamefont
  {Gunacker}, \citenamefont {Wallerberger}, \citenamefont {Ribic},
  \citenamefont {Hausoel}, \citenamefont {Sangiovanni},\ and\ \citenamefont
  {Held}}]{Gunacker16}%
  \BibitemOpen
  \bibfield  {author} {\bibinfo {author} {\bibfnamefont {P.}~\bibnamefont
  {Gunacker}}, \bibinfo {author} {\bibfnamefont {M.}~\bibnamefont
  {Wallerberger}}, \bibinfo {author} {\bibfnamefont {T.}~\bibnamefont {Ribic}},
  \bibinfo {author} {\bibfnamefont {A.}~\bibnamefont {Hausoel}}, \bibinfo
  {author} {\bibfnamefont {G.}~\bibnamefont {Sangiovanni}}, \ and\ \bibinfo
  {author} {\bibfnamefont {K.}~\bibnamefont {Held}},\ }\href {\doibase
  10.1103/PhysRevB.94.125153} {\bibfield  {journal} {\bibinfo  {journal} {Phys.
  Rev. B}\ }\textbf {\bibinfo {volume} {94}},\ \bibinfo {pages} {125153}
  (\bibinfo {year} {2016})}\BibitemShut {NoStop}%
\bibitem [{\citenamefont {Boehnke}\ \emph {et~al.}(2018)\citenamefont
  {Boehnke}, \citenamefont {Werner},\ and\ \citenamefont
  {Lechermann}}]{Boehnke18}%
  \BibitemOpen
  \bibfield  {author} {\bibinfo {author} {\bibfnamefont {L.}~\bibnamefont
  {Boehnke}}, \bibinfo {author} {\bibfnamefont {P.}~\bibnamefont {Werner}}, \
  and\ \bibinfo {author} {\bibfnamefont {F.}~\bibnamefont {Lechermann}},\
  }\href {\doibase 10.1209/0295-5075/122/57001} {\bibfield  {journal} {\bibinfo
   {journal} {{EPL} (Europhysics Letters)}\ }\textbf {\bibinfo {volume}
  {122}},\ \bibinfo {pages} {57001} (\bibinfo {year} {2018})}\BibitemShut
  {NoStop}%
\bibitem [{\citenamefont {Tanaka}(2019)}]{Tanaka18}%
  \BibitemOpen
  \bibfield  {author} {\bibinfo {author} {\bibfnamefont {A.}~\bibnamefont
  {Tanaka}},\ }\href {\doibase 10.1103/PhysRevB.99.205133} {\bibfield
  {journal} {\bibinfo  {journal} {Phys. Rev. B}\ }\textbf {\bibinfo {volume}
  {99}},\ \bibinfo {pages} {205133} (\bibinfo {year} {2019})}\BibitemShut
  {NoStop}%
\bibitem [{\citenamefont {Hafermann}\ \emph {et~al.}(2014)\citenamefont
  {Hafermann}, \citenamefont {van Loon}, \citenamefont {Katsnelson},
  \citenamefont {Lichtenstein},\ and\ \citenamefont
  {Parcollet}}]{Hafermann14-2}%
  \BibitemOpen
  \bibfield  {author} {\bibinfo {author} {\bibfnamefont {H.}~\bibnamefont
  {Hafermann}}, \bibinfo {author} {\bibfnamefont {E.~G. C.~P.}\ \bibnamefont
  {van Loon}}, \bibinfo {author} {\bibfnamefont {M.~I.}\ \bibnamefont
  {Katsnelson}}, \bibinfo {author} {\bibfnamefont {A.~I.}\ \bibnamefont
  {Lichtenstein}}, \ and\ \bibinfo {author} {\bibfnamefont {O.}~\bibnamefont
  {Parcollet}},\ }\href {\doibase 10.1103/PhysRevB.90.235105} {\bibfield
  {journal} {\bibinfo  {journal} {Phys. Rev. B}\ }\textbf {\bibinfo {volume}
  {90}},\ \bibinfo {pages} {235105} (\bibinfo {year} {2014})}\BibitemShut
  {NoStop}%
\bibitem [{\citenamefont {Kune\ifmmode~\check{s}\else
  \v{s}\fi{}}(2011)}]{Kunes11}%
  \BibitemOpen
  \bibfield  {author} {\bibinfo {author} {\bibfnamefont {J.}~\bibnamefont
  {Kune\ifmmode~\check{s}\else \v{s}\fi{}}},\ }\href {\doibase
  10.1103/PhysRevB.83.085102} {\bibfield  {journal} {\bibinfo  {journal} {Phys.
  Rev. B}\ }\textbf {\bibinfo {volume} {83}},\ \bibinfo {pages} {085102}
  (\bibinfo {year} {2011})}\BibitemShut {NoStop}%
\bibitem [{\citenamefont {Wentzell}\ \emph {et~al.}(2016)\citenamefont
  {Wentzell}, \citenamefont {Li}, \citenamefont {Tagliavini}, \citenamefont
  {Taranto}, \citenamefont {Rohringer}, \citenamefont {Held}, \citenamefont
  {Toschi},\ and\ \citenamefont {Andergassen}}]{Wentzell16}%
  \BibitemOpen
  \bibfield  {author} {\bibinfo {author} {\bibfnamefont {N.}~\bibnamefont
  {Wentzell}}, \bibinfo {author} {\bibfnamefont {G.}~\bibnamefont {Li}},
  \bibinfo {author} {\bibfnamefont {A.}~\bibnamefont {Tagliavini}}, \bibinfo
  {author} {\bibfnamefont {C.}~\bibnamefont {Taranto}}, \bibinfo {author}
  {\bibfnamefont {G.}~\bibnamefont {Rohringer}}, \bibinfo {author}
  {\bibfnamefont {K.}~\bibnamefont {Held}}, \bibinfo {author} {\bibfnamefont
  {A.}~\bibnamefont {Toschi}}, \ and\ \bibinfo {author} {\bibfnamefont
  {S.}~\bibnamefont {Andergassen}},\ }\href@noop {} {\enquote {\bibinfo {title}
  {High-frequency asymptotics of the vertex function},}\ } (\bibinfo {year}
  {2016}),\ \Eprint {http://arxiv.org/abs/arXiv:1610.06520} {arXiv:1610.06520}
  \BibitemShut {NoStop}%
\bibitem [{\citenamefont {Kaufmann}\ \emph {et~al.}(2017)\citenamefont
  {Kaufmann}, \citenamefont {Gunacker},\ and\ \citenamefont
  {Held}}]{Kaufmann17}%
  \BibitemOpen
  \bibfield  {author} {\bibinfo {author} {\bibfnamefont {J.}~\bibnamefont
  {Kaufmann}}, \bibinfo {author} {\bibfnamefont {P.}~\bibnamefont {Gunacker}},
  \ and\ \bibinfo {author} {\bibfnamefont {K.}~\bibnamefont {Held}},\ }\href
  {\doibase 10.1103/PhysRevB.96.035114} {\bibfield  {journal} {\bibinfo
  {journal} {Phys. Rev. B}\ }\textbf {\bibinfo {volume} {96}},\ \bibinfo
  {pages} {035114} (\bibinfo {year} {2017})}\BibitemShut {NoStop}%
\bibitem [{\citenamefont {Tagliavini}\ \emph {et~al.}(2018)\citenamefont
  {Tagliavini}, \citenamefont {Hummel}, \citenamefont {Wentzell}, \citenamefont
  {Andergassen}, \citenamefont {Toschi},\ and\ \citenamefont
  {Rohringer}}]{Tagliavini18}%
  \BibitemOpen
  \bibfield  {author} {\bibinfo {author} {\bibfnamefont {A.}~\bibnamefont
  {Tagliavini}}, \bibinfo {author} {\bibfnamefont {S.}~\bibnamefont {Hummel}},
  \bibinfo {author} {\bibfnamefont {N.}~\bibnamefont {Wentzell}}, \bibinfo
  {author} {\bibfnamefont {S.}~\bibnamefont {Andergassen}}, \bibinfo {author}
  {\bibfnamefont {A.}~\bibnamefont {Toschi}}, \ and\ \bibinfo {author}
  {\bibfnamefont {G.}~\bibnamefont {Rohringer}},\ }\href {\doibase
  10.1103/PhysRevB.97.235140} {\bibfield  {journal} {\bibinfo  {journal} {Phys.
  Rev. B}\ }\textbf {\bibinfo {volume} {97}},\ \bibinfo {pages} {235140}
  (\bibinfo {year} {2018})}\BibitemShut {NoStop}%
\bibitem [{\citenamefont {Pruschke}\ \emph {et~al.}(1996)\citenamefont
  {Pruschke}, \citenamefont {Qin}, \citenamefont {Obermeier},\ and\
  \citenamefont {Keller}}]{Pruschke96}%
  \BibitemOpen
  \bibfield  {author} {\bibinfo {author} {\bibfnamefont {T.}~\bibnamefont
  {Pruschke}}, \bibinfo {author} {\bibfnamefont {Q.}~\bibnamefont {Qin}},
  \bibinfo {author} {\bibfnamefont {T.}~\bibnamefont {Obermeier}}, \ and\
  \bibinfo {author} {\bibfnamefont {J.}~\bibnamefont {Keller}},\ }\href
  {http://stacks.iop.org/0953-8984/8/i=18/a=009} {\bibfield  {journal}
  {\bibinfo  {journal} {Journal of Physics: Condensed Matter}\ }\textbf
  {\bibinfo {volume} {8}},\ \bibinfo {pages} {3161} (\bibinfo {year}
  {1996})}\BibitemShut {NoStop}%
\bibitem [{\citenamefont {Rubtsov}\ \emph {et~al.}(2012)\citenamefont
  {Rubtsov}, \citenamefont {Katsnelson},\ and\ \citenamefont
  {Lichtenstein}}]{Rubtsov12}%
  \BibitemOpen
  \bibfield  {author} {\bibinfo {author} {\bibfnamefont {A.~N.}\ \bibnamefont
  {Rubtsov}}, \bibinfo {author} {\bibfnamefont {M.~I.}\ \bibnamefont
  {Katsnelson}}, \ and\ \bibinfo {author} {\bibfnamefont {A.~I.}\ \bibnamefont
  {Lichtenstein}},\ }\href {\doibase 10.1016/j.aop.2012.01.002} {\bibfield
  {journal} {\bibinfo  {journal} {Annals of Physics}\ }\textbf {\bibinfo
  {volume} {327}},\ \bibinfo {pages} {1320} (\bibinfo {year}
  {2012})}\BibitemShut {NoStop}%
\bibitem [{\citenamefont {van Loon}\ \emph {et~al.}(2016)\citenamefont {van
  Loon}, \citenamefont {Krien}, \citenamefont {Hafermann}, \citenamefont
  {Stepanov}, \citenamefont {Lichtenstein},\ and\ \citenamefont
  {Katsnelson}}]{vanLoon16}%
  \BibitemOpen
  \bibfield  {author} {\bibinfo {author} {\bibfnamefont {E.~G. C.~P.}\
  \bibnamefont {van Loon}}, \bibinfo {author} {\bibfnamefont {F.}~\bibnamefont
  {Krien}}, \bibinfo {author} {\bibfnamefont {H.}~\bibnamefont {Hafermann}},
  \bibinfo {author} {\bibfnamefont {E.~A.}\ \bibnamefont {Stepanov}}, \bibinfo
  {author} {\bibfnamefont {A.~I.}\ \bibnamefont {Lichtenstein}}, \ and\
  \bibinfo {author} {\bibfnamefont {M.~I.}\ \bibnamefont {Katsnelson}},\ }\href
  {\doibase 10.1103/PhysRevB.93.155162} {\bibfield  {journal} {\bibinfo
  {journal} {Phys. Rev. B}\ }\textbf {\bibinfo {volume} {93}},\ \bibinfo
  {pages} {155162} (\bibinfo {year} {2016})}\BibitemShut {NoStop}%
\bibitem [{\citenamefont {Stepanov}\ \emph
  {et~al.}(2016{\natexlab{a}})\citenamefont {Stepanov}, \citenamefont {Huber},
  \citenamefont {van Loon}, \citenamefont {Lichtenstein},\ and\ \citenamefont
  {Katsnelson}}]{Stepanov16-2}%
  \BibitemOpen
  \bibfield  {author} {\bibinfo {author} {\bibfnamefont {E.~A.}\ \bibnamefont
  {Stepanov}}, \bibinfo {author} {\bibfnamefont {A.}~\bibnamefont {Huber}},
  \bibinfo {author} {\bibfnamefont {E.~G. C.~P.}\ \bibnamefont {van Loon}},
  \bibinfo {author} {\bibfnamefont {A.~I.}\ \bibnamefont {Lichtenstein}}, \
  and\ \bibinfo {author} {\bibfnamefont {M.~I.}\ \bibnamefont {Katsnelson}},\
  }\href {\doibase 10.1103/PhysRevB.94.205110} {\bibfield  {journal} {\bibinfo
  {journal} {Phys. Rev. B}\ }\textbf {\bibinfo {volume} {94}},\ \bibinfo
  {pages} {205110} (\bibinfo {year} {2016}{\natexlab{a}})}\BibitemShut
  {NoStop}%
\bibitem [{\citenamefont {Hedin}(1965)}]{Hedin65}%
  \BibitemOpen
  \bibfield  {author} {\bibinfo {author} {\bibfnamefont {L.}~\bibnamefont
  {Hedin}},\ }\href {\doibase 10.1103/PhysRev.139.A796} {\bibfield  {journal}
  {\bibinfo  {journal} {Phys. Rev.}\ }\textbf {\bibinfo {volume} {139}},\
  \bibinfo {pages} {A796} (\bibinfo {year} {1965})}\BibitemShut {NoStop}%
\bibitem [{\citenamefont {Ayral}\ and\ \citenamefont
  {Parcollet}(2016)}]{Ayral15-2}%
  \BibitemOpen
  \bibfield  {author} {\bibinfo {author} {\bibfnamefont {T.}~\bibnamefont
  {Ayral}}\ and\ \bibinfo {author} {\bibfnamefont {O.}~\bibnamefont
  {Parcollet}},\ }\href {\doibase 10.1103/PhysRevB.93.235124} {\bibfield
  {journal} {\bibinfo  {journal} {Phys. Rev. B}\ }\textbf {\bibinfo {volume}
  {93}},\ \bibinfo {pages} {235124} (\bibinfo {year} {2016})}\BibitemShut
  {NoStop}%
\bibitem [{\citenamefont {Otsuki}\ \emph {et~al.}(2014)\citenamefont {Otsuki},
  \citenamefont {Hafermann},\ and\ \citenamefont {Lichtenstein}}]{Otsuki14}%
  \BibitemOpen
  \bibfield  {author} {\bibinfo {author} {\bibfnamefont {J.}~\bibnamefont
  {Otsuki}}, \bibinfo {author} {\bibfnamefont {H.}~\bibnamefont {Hafermann}}, \
  and\ \bibinfo {author} {\bibfnamefont {A.~I.}\ \bibnamefont {Lichtenstein}},\
  }\href {\doibase 10.1103/PhysRevB.90.235132} {\bibfield  {journal} {\bibinfo
  {journal} {Phys. Rev. B}\ }\textbf {\bibinfo {volume} {90}},\ \bibinfo
  {pages} {235132} (\bibinfo {year} {2014})}\BibitemShut {NoStop}%
\bibitem [{\citenamefont {Katanin}\ \emph {et~al.}(2009)\citenamefont
  {Katanin}, \citenamefont {Toschi},\ and\ \citenamefont {Held}}]{Katanin09}%
  \BibitemOpen
  \bibfield  {author} {\bibinfo {author} {\bibfnamefont {A.~A.}\ \bibnamefont
  {Katanin}}, \bibinfo {author} {\bibfnamefont {A.}~\bibnamefont {Toschi}}, \
  and\ \bibinfo {author} {\bibfnamefont {K.}~\bibnamefont {Held}},\ }\href
  {\doibase 10.1103/PhysRevB.80.075104} {\bibfield  {journal} {\bibinfo
  {journal} {Phys. Rev. B}\ }\textbf {\bibinfo {volume} {80}},\ \bibinfo
  {pages} {075104} (\bibinfo {year} {2009})}\BibitemShut {NoStop}%
\bibitem [{\citenamefont {Rohringer}\ \emph {et~al.}(2011)\citenamefont
  {Rohringer}, \citenamefont {Toschi}, \citenamefont {Katanin},\ and\
  \citenamefont {Held}}]{Rohringer11}%
  \BibitemOpen
  \bibfield  {author} {\bibinfo {author} {\bibfnamefont {G.}~\bibnamefont
  {Rohringer}}, \bibinfo {author} {\bibfnamefont {A.}~\bibnamefont {Toschi}},
  \bibinfo {author} {\bibfnamefont {A.}~\bibnamefont {Katanin}}, \ and\
  \bibinfo {author} {\bibfnamefont {K.}~\bibnamefont {Held}},\ }\href {\doibase
  10.1103/PhysRevLett.107.256402} {\bibfield  {journal} {\bibinfo  {journal}
  {Phys. Rev. Lett.}\ }\textbf {\bibinfo {volume} {107}},\ \bibinfo {pages}
  {256402} (\bibinfo {year} {2011})}\BibitemShut {NoStop}%
\bibitem [{\citenamefont {Hirschmeier}\ \emph {et~al.}(2015)\citenamefont
  {Hirschmeier}, \citenamefont {Hafermann}, \citenamefont {Gull}, \citenamefont
  {Lichtenstein},\ and\ \citenamefont {Antipov}}]{Hirschmeier15}%
  \BibitemOpen
  \bibfield  {author} {\bibinfo {author} {\bibfnamefont {D.}~\bibnamefont
  {Hirschmeier}}, \bibinfo {author} {\bibfnamefont {H.}~\bibnamefont
  {Hafermann}}, \bibinfo {author} {\bibfnamefont {E.}~\bibnamefont {Gull}},
  \bibinfo {author} {\bibfnamefont {A.~I.}\ \bibnamefont {Lichtenstein}}, \
  and\ \bibinfo {author} {\bibfnamefont {A.~E.}\ \bibnamefont {Antipov}},\
  }\href {\doibase 10.1103/PhysRevB.92.144409} {\bibfield  {journal} {\bibinfo
  {journal} {Phys. Rev. B}\ }\textbf {\bibinfo {volume} {92}},\ \bibinfo
  {pages} {144409} (\bibinfo {year} {2015})}\BibitemShut {NoStop}%
\bibitem [{\citenamefont {Rohringer}\ and\ \citenamefont
  {Toschi}(2016)}]{Rohringer16}%
  \BibitemOpen
  \bibfield  {author} {\bibinfo {author} {\bibfnamefont {G.}~\bibnamefont
  {Rohringer}}\ and\ \bibinfo {author} {\bibfnamefont {A.}~\bibnamefont
  {Toschi}},\ }\href {\doibase 10.1103/PhysRevB.94.125144} {\bibfield
  {journal} {\bibinfo  {journal} {Phys. Rev. B}\ }\textbf {\bibinfo {volume}
  {94}},\ \bibinfo {pages} {125144} (\bibinfo {year} {2016})}\BibitemShut
  {NoStop}%
\bibitem [{\citenamefont {{Y.M. Vilk}}\ and\ \citenamefont {{A.-M.S.
  Tremblay}}(1997)}]{Vilk97}%
  \BibitemOpen
  \bibfield  {author} {\bibinfo {author} {\bibnamefont {{Y.M. Vilk}}}\ and\
  \bibinfo {author} {\bibnamefont {{A.-M.S. Tremblay}}},\ }\href {\doibase
  10.1051/jp1:1997135} {\bibfield  {journal} {\bibinfo  {journal} {J. Phys. I
  France}\ }\textbf {\bibinfo {volume} {7}},\ \bibinfo {pages} {1309} (\bibinfo
  {year} {1997})}\BibitemShut {NoStop}%
\bibitem [{\citenamefont {van Loon}\ \emph
  {et~al.}(2018{\natexlab{a}})\citenamefont {van Loon}, \citenamefont {Krien},
  \citenamefont {Hafermann}, \citenamefont {Lichtenstein},\ and\ \citenamefont
  {Katsnelson}}]{vanLoon18}%
  \BibitemOpen
  \bibfield  {author} {\bibinfo {author} {\bibfnamefont {E.~G. C.~P.}\
  \bibnamefont {van Loon}}, \bibinfo {author} {\bibfnamefont {F.}~\bibnamefont
  {Krien}}, \bibinfo {author} {\bibfnamefont {H.}~\bibnamefont {Hafermann}},
  \bibinfo {author} {\bibfnamefont {A.~I.}\ \bibnamefont {Lichtenstein}}, \
  and\ \bibinfo {author} {\bibfnamefont {M.~I.}\ \bibnamefont {Katsnelson}},\
  }\href {\doibase 10.1103/PhysRevB.98.205148} {\bibfield  {journal} {\bibinfo
  {journal} {Phys. Rev. B}\ }\textbf {\bibinfo {volume} {98}},\ \bibinfo
  {pages} {205148} (\bibinfo {year} {2018}{\natexlab{a}})}\BibitemShut
  {NoStop}%
\bibitem [{\citenamefont {Hertz}\ and\ \citenamefont
  {Edwards}(1973)}]{Hertz73}%
  \BibitemOpen
  \bibfield  {author} {\bibinfo {author} {\bibfnamefont {J.~A.}\ \bibnamefont
  {Hertz}}\ and\ \bibinfo {author} {\bibfnamefont {D.~M.}\ \bibnamefont
  {Edwards}},\ }\href {http://stacks.iop.org/0305-4608/3/i=12/a=018} {\bibfield
   {journal} {\bibinfo  {journal} {Journal of Physics F: Metal Physics}\
  }\textbf {\bibinfo {volume} {3}},\ \bibinfo {pages} {2174} (\bibinfo {year}
  {1973})}\BibitemShut {NoStop}%
\bibitem [{Note1()}]{Note1}%
  \BibitemOpen
  \bibinfo {note} {The polarization $\pi $ can indeed be interpreted as the
  `self-energy' of the screened interaction $w$, analogous to the Dyson
  equation $g=g^0/(1-g^0\Sigma )$, and the bare interaction $U$ assumes the
  role of the bare Green's function $g^0$. On the other hand, $U$ also
  corresponds to the \protect \textit {two-particle} self-energy of the RPA
  approximation~\cite {Mahan00,Vilk96}, one may therefore refer to the diagrams
  in Fig.~\ref {fig:irrvertices} c) as `RPA-like'.}\BibitemShut {Stop}%
\bibitem [{\citenamefont {Held}\ \emph {et~al.}(2011)\citenamefont {Held},
  \citenamefont {Taranto}, \citenamefont {Rohringer},\ and\ \citenamefont
  {Toschi}}]{Held11}%
  \BibitemOpen
  \bibfield  {author} {\bibinfo {author} {\bibfnamefont {K.}~\bibnamefont
  {Held}}, \bibinfo {author} {\bibfnamefont {C.}~\bibnamefont {Taranto}},
  \bibinfo {author} {\bibfnamefont {G.}~\bibnamefont {Rohringer}}, \ and\
  \bibinfo {author} {\bibfnamefont {A.}~\bibnamefont {Toschi}},\ }\href@noop {}
  {\enquote {\bibinfo {title} {Hedin equations, {GW}, {GW+DMFT}, and all
  that},}\ } (\bibinfo {year} {2011}),\ \Eprint
  {http://arxiv.org/abs/arXiv:1109.3972} {arXiv:1109.3972} \BibitemShut
  {NoStop}%
\bibitem [{\citenamefont {Otsuki}\ \emph {et~al.}(2019)\citenamefont {Otsuki},
  \citenamefont {Yoshimi}, \citenamefont {Shinaoka},\ and\ \citenamefont
  {Nomura}}]{Otsuki19}%
  \BibitemOpen
  \bibfield  {author} {\bibinfo {author} {\bibfnamefont {J.}~\bibnamefont
  {Otsuki}}, \bibinfo {author} {\bibfnamefont {K.}~\bibnamefont {Yoshimi}},
  \bibinfo {author} {\bibfnamefont {H.}~\bibnamefont {Shinaoka}}, \ and\
  \bibinfo {author} {\bibfnamefont {Y.}~\bibnamefont {Nomura}},\ }\href
  {\doibase 10.1103/PhysRevB.99.165134} {\bibfield  {journal} {\bibinfo
  {journal} {Phys. Rev. B}\ }\textbf {\bibinfo {volume} {99}},\ \bibinfo
  {pages} {165134} (\bibinfo {year} {2019})}\BibitemShut {NoStop}%
\bibitem [{\citenamefont {Krien}\ \emph {et~al.}(2017)\citenamefont {Krien},
  \citenamefont {van Loon}, \citenamefont {Hafermann}, \citenamefont {Otsuki},
  \citenamefont {Katsnelson},\ and\ \citenamefont {Lichtenstein}}]{Krien17}%
  \BibitemOpen
  \bibfield  {author} {\bibinfo {author} {\bibfnamefont {F.}~\bibnamefont
  {Krien}}, \bibinfo {author} {\bibfnamefont {E.~G. C.~P.}\ \bibnamefont {van
  Loon}}, \bibinfo {author} {\bibfnamefont {H.}~\bibnamefont {Hafermann}},
  \bibinfo {author} {\bibfnamefont {J.}~\bibnamefont {Otsuki}}, \bibinfo
  {author} {\bibfnamefont {M.~I.}\ \bibnamefont {Katsnelson}}, \ and\ \bibinfo
  {author} {\bibfnamefont {A.~I.}\ \bibnamefont {Lichtenstein}},\ }\href
  {\doibase 10.1103/PhysRevB.96.075155} {\bibfield  {journal} {\bibinfo
  {journal} {Phys. Rev. B}\ }\textbf {\bibinfo {volume} {96}},\ \bibinfo
  {pages} {075155} (\bibinfo {year} {2017})}\BibitemShut {NoStop}%
\bibitem [{Note2()}]{Note2}%
  \BibitemOpen
  \bibinfo {note} {This does not directly imply satisfaction of the
  Mermin-Wagner theorem, because it has to be shown in practice that a solution
  $\protect \mathcal {U}_\omega $ exists that satisfies Eq.~\protect \textup
  {\hbox {\mathsurround \z@ \protect \normalfont (\ignorespaces \ref
  {eq:tpsc}\unskip \@@italiccorr )}}.}\BibitemShut {Stop}%
\bibitem [{Note3()}]{Note3}%
  \BibitemOpen
  \bibinfo {note} {Despite the ungapped spectrum the Ward identity is
  nevertheless violated, because due to the correction $\protect \mathcal {U}$
  the static homogeneous limit of $\protect \mathcal {X}$ is inconsistent with
  the one-particle level of the DMFT approximation~\cite
  {Krien18}.}\BibitemShut {Stop}%
\bibitem [{\citenamefont {Dar\'e}\ \emph {et~al.}(1996)\citenamefont {Dar\'e},
  \citenamefont {Vilk},\ and\ \citenamefont {Tremblay}}]{Dare96}%
  \BibitemOpen
  \bibfield  {author} {\bibinfo {author} {\bibfnamefont {A.-M.}\ \bibnamefont
  {Dar\'e}}, \bibinfo {author} {\bibfnamefont {Y.~M.}\ \bibnamefont {Vilk}}, \
  and\ \bibinfo {author} {\bibfnamefont {A.~M.~S.}\ \bibnamefont {Tremblay}},\
  }\href {\doibase 10.1103/PhysRevB.53.14236} {\bibfield  {journal} {\bibinfo
  {journal} {Phys. Rev. B}\ }\textbf {\bibinfo {volume} {53}},\ \bibinfo
  {pages} {14236} (\bibinfo {year} {1996})}\BibitemShut {NoStop}%
\bibitem [{\citenamefont {Bauer}\ \emph {et~al.}(2011)\citenamefont {Bauer},
  \citenamefont {Carr}, \citenamefont {Evertz}, \citenamefont {Feiguin},
  \citenamefont {Freire}, \citenamefont {Fuchs}, \citenamefont {Gamper},
  \citenamefont {Gukelberger}, \citenamefont {Gull}, \citenamefont {Guertler},
  \citenamefont {Hehn}, \citenamefont {Igarashi}, \citenamefont {Isakov},
  \citenamefont {Koop}, \citenamefont {Ma}, \citenamefont {Mates},
  \citenamefont {Matsuo}, \citenamefont {Parcollet}, \citenamefont
  {Pawłowski}, \citenamefont {Picon}, \citenamefont {Pollet}, \citenamefont
  {Santos}, \citenamefont {Scarola}, \citenamefont {Schollw\"ock},
  \citenamefont {Silva}, \citenamefont {Surer}, \citenamefont {Todo},
  \citenamefont {Trebst}, \citenamefont {Troyer}, \citenamefont {Wall},
  \citenamefont {Werner},\ and\ \citenamefont {Wessel}}]{ALPS2}%
  \BibitemOpen
  \bibfield  {author} {\bibinfo {author} {\bibfnamefont {B.}~\bibnamefont
  {Bauer}}, \bibinfo {author} {\bibfnamefont {L.~D.}\ \bibnamefont {Carr}},
  \bibinfo {author} {\bibfnamefont {H.~G.}\ \bibnamefont {Evertz}}, \bibinfo
  {author} {\bibfnamefont {A.}~\bibnamefont {Feiguin}}, \bibinfo {author}
  {\bibfnamefont {J.}~\bibnamefont {Freire}}, \bibinfo {author} {\bibfnamefont
  {S.}~\bibnamefont {Fuchs}}, \bibinfo {author} {\bibfnamefont
  {L.}~\bibnamefont {Gamper}}, \bibinfo {author} {\bibfnamefont
  {J.}~\bibnamefont {Gukelberger}}, \bibinfo {author} {\bibfnamefont
  {E.}~\bibnamefont {Gull}}, \bibinfo {author} {\bibfnamefont {S.}~\bibnamefont
  {Guertler}}, \bibinfo {author} {\bibfnamefont {A.}~\bibnamefont {Hehn}},
  \bibinfo {author} {\bibfnamefont {R.}~\bibnamefont {Igarashi}}, \bibinfo
  {author} {\bibfnamefont {S.~V.}\ \bibnamefont {Isakov}}, \bibinfo {author}
  {\bibfnamefont {D.}~\bibnamefont {Koop}}, \bibinfo {author} {\bibfnamefont
  {P.~N.}\ \bibnamefont {Ma}}, \bibinfo {author} {\bibfnamefont
  {P.}~\bibnamefont {Mates}}, \bibinfo {author} {\bibfnamefont
  {H.}~\bibnamefont {Matsuo}}, \bibinfo {author} {\bibfnamefont
  {O.}~\bibnamefont {Parcollet}}, \bibinfo {author} {\bibfnamefont
  {G.}~\bibnamefont {Pawłowski}}, \bibinfo {author} {\bibfnamefont {J.~D.}\
  \bibnamefont {Picon}}, \bibinfo {author} {\bibfnamefont {L.}~\bibnamefont
  {Pollet}}, \bibinfo {author} {\bibfnamefont {E.}~\bibnamefont {Santos}},
  \bibinfo {author} {\bibfnamefont {V.~W.}\ \bibnamefont {Scarola}}, \bibinfo
  {author} {\bibfnamefont {U.}~\bibnamefont {Schollw\"ock}}, \bibinfo {author}
  {\bibfnamefont {C.}~\bibnamefont {Silva}}, \bibinfo {author} {\bibfnamefont
  {B.}~\bibnamefont {Surer}}, \bibinfo {author} {\bibfnamefont
  {S.}~\bibnamefont {Todo}}, \bibinfo {author} {\bibfnamefont {S.}~\bibnamefont
  {Trebst}}, \bibinfo {author} {\bibfnamefont {M.}~\bibnamefont {Troyer}},
  \bibinfo {author} {\bibfnamefont {M.~L.}\ \bibnamefont {Wall}}, \bibinfo
  {author} {\bibfnamefont {P.}~\bibnamefont {Werner}}, \ and\ \bibinfo {author}
  {\bibfnamefont {S.}~\bibnamefont {Wessel}},\ }\href {\doibase
  10.1088/1742-5468/2011/05/P05001} {\bibfield  {journal} {\bibinfo  {journal}
  {Journal of Statistical Mechanics: Theory and Experiment}\ }\textbf {\bibinfo
  {volume} {2011}},\ \bibinfo {pages} {P05001} (\bibinfo {year}
  {2011})}\BibitemShut {NoStop}%
\bibitem [{\citenamefont {van Loon}\ \emph
  {et~al.}(2014{\natexlab{b}})\citenamefont {van Loon}, \citenamefont
  {Lichtenstein}, \citenamefont {Katsnelson}, \citenamefont {Parcollet},\ and\
  \citenamefont {Hafermann}}]{vanLoon14}%
  \BibitemOpen
  \bibfield  {author} {\bibinfo {author} {\bibfnamefont {E.~G. C.~P.}\
  \bibnamefont {van Loon}}, \bibinfo {author} {\bibfnamefont {A.~I.}\
  \bibnamefont {Lichtenstein}}, \bibinfo {author} {\bibfnamefont {M.~I.}\
  \bibnamefont {Katsnelson}}, \bibinfo {author} {\bibfnamefont
  {O.}~\bibnamefont {Parcollet}}, \ and\ \bibinfo {author} {\bibfnamefont
  {H.}~\bibnamefont {Hafermann}},\ }\href {\doibase 10.1103/PhysRevB.90.235135}
  {\bibfield  {journal} {\bibinfo  {journal} {Phys. Rev. B}\ }\textbf {\bibinfo
  {volume} {90}},\ \bibinfo {pages} {235135} (\bibinfo {year}
  {2014}{\natexlab{b}})}\BibitemShut {NoStop}%
\bibitem [{\citenamefont {Stepanov}\ \emph
  {et~al.}(2016{\natexlab{b}})\citenamefont {Stepanov}, \citenamefont {van
  Loon}, \citenamefont {Katanin}, \citenamefont {Lichtenstein}, \citenamefont
  {Katsnelson},\ and\ \citenamefont {Rubtsov}}]{Stepanov16}%
  \BibitemOpen
  \bibfield  {author} {\bibinfo {author} {\bibfnamefont {E.~A.}\ \bibnamefont
  {Stepanov}}, \bibinfo {author} {\bibfnamefont {E.~G. C.~P.}\ \bibnamefont
  {van Loon}}, \bibinfo {author} {\bibfnamefont {A.~A.}\ \bibnamefont
  {Katanin}}, \bibinfo {author} {\bibfnamefont {A.~I.}\ \bibnamefont
  {Lichtenstein}}, \bibinfo {author} {\bibfnamefont {M.~I.}\ \bibnamefont
  {Katsnelson}}, \ and\ \bibinfo {author} {\bibfnamefont {A.~N.}\ \bibnamefont
  {Rubtsov}},\ }\href {\doibase 10.1103/PhysRevB.93.045107} {\bibfield
  {journal} {\bibinfo  {journal} {Phys. Rev. B}\ }\textbf {\bibinfo {volume}
  {93}},\ \bibinfo {pages} {045107} (\bibinfo {year}
  {2016}{\natexlab{b}})}\BibitemShut {NoStop}%
\bibitem [{Note4()}]{Note4}%
  \BibitemOpen
  \bibinfo {note} {The similar results are a consequence of similar
  self-consistency conditions. The Moriya-$\lambda $ is fixed by the local sum
  rules~\protect \textup {\hbox {\mathsurround \z@ \protect \normalfont
  (\ignorespaces \ref {eq:locch}\unskip \@@italiccorr )}} and~\protect \textup
  {\hbox {\mathsurround \z@ \protect \normalfont (\ignorespaces \ref
  {eq:locsp}\unskip \@@italiccorr )}}, whose left-hand-sides are in general
  dominated by the static term $\omega =0$ near a phase transition, in this
  case $\protect \mathcal {U}^\alpha (\omega =0)\approx \lambda ^\alpha
  _\protect \text {Moriya}$.}\BibitemShut {Stop}%
\bibitem [{\citenamefont {Rubtsov}\ \emph {et~al.}(2008)\citenamefont
  {Rubtsov}, \citenamefont {Katsnelson},\ and\ \citenamefont
  {Lichtenstein}}]{Rubtsov08}%
  \BibitemOpen
  \bibfield  {author} {\bibinfo {author} {\bibfnamefont {A.~N.}\ \bibnamefont
  {Rubtsov}}, \bibinfo {author} {\bibfnamefont {M.~I.}\ \bibnamefont
  {Katsnelson}}, \ and\ \bibinfo {author} {\bibfnamefont {A.~I.}\ \bibnamefont
  {Lichtenstein}},\ }\href {\doibase 10.1103/PhysRevB.77.033101} {\bibfield
  {journal} {\bibinfo  {journal} {Phys. Rev. B}\ }\textbf {\bibinfo {volume}
  {77}},\ \bibinfo {pages} {033101} (\bibinfo {year} {2008})}\BibitemShut
  {NoStop}%
\bibitem [{\citenamefont {Vilk}\ and\ \citenamefont {Tremblay}(1996)}]{Vilk96}%
  \BibitemOpen
  \bibfield  {author} {\bibinfo {author} {\bibfnamefont {Y.~M.}\ \bibnamefont
  {Vilk}}\ and\ \bibinfo {author} {\bibfnamefont {A.-M.~S.}\ \bibnamefont
  {Tremblay}},\ }\href {http://stacks.iop.org/0295-5075/33/i=2/a=159}
  {\bibfield  {journal} {\bibinfo  {journal} {EPL (Europhysics Letters)}\
  }\textbf {\bibinfo {volume} {33}},\ \bibinfo {pages} {159} (\bibinfo {year}
  {1996})}\BibitemShut {NoStop}%
\bibitem [{\citenamefont {Sch\"afer}\ \emph {et~al.}(2015)\citenamefont
  {Sch\"afer}, \citenamefont {Geles}, \citenamefont {Rost}, \citenamefont
  {Rohringer}, \citenamefont {Arrigoni}, \citenamefont {Held}, \citenamefont
  {Bl\"umer}, \citenamefont {Aichhorn},\ and\ \citenamefont
  {Toschi}}]{Schaefer15}%
  \BibitemOpen
  \bibfield  {author} {\bibinfo {author} {\bibfnamefont {T.}~\bibnamefont
  {Sch\"afer}}, \bibinfo {author} {\bibfnamefont {F.}~\bibnamefont {Geles}},
  \bibinfo {author} {\bibfnamefont {D.}~\bibnamefont {Rost}}, \bibinfo {author}
  {\bibfnamefont {G.}~\bibnamefont {Rohringer}}, \bibinfo {author}
  {\bibfnamefont {E.}~\bibnamefont {Arrigoni}}, \bibinfo {author}
  {\bibfnamefont {K.}~\bibnamefont {Held}}, \bibinfo {author} {\bibfnamefont
  {N.}~\bibnamefont {Bl\"umer}}, \bibinfo {author} {\bibfnamefont
  {M.}~\bibnamefont {Aichhorn}}, \ and\ \bibinfo {author} {\bibfnamefont
  {A.}~\bibnamefont {Toschi}},\ }\href {\doibase 10.1103/PhysRevB.91.125109}
  {\bibfield  {journal} {\bibinfo  {journal} {Phys. Rev. B}\ }\textbf {\bibinfo
  {volume} {91}},\ \bibinfo {pages} {125109} (\bibinfo {year}
  {2015})}\BibitemShut {NoStop}%
\bibitem [{\citenamefont {van Loon}\ \emph
  {et~al.}(2018{\natexlab{b}})\citenamefont {van Loon}, \citenamefont
  {Hafermann},\ and\ \citenamefont {Katsnelson}}]{vanLoon18-2}%
  \BibitemOpen
  \bibfield  {author} {\bibinfo {author} {\bibfnamefont {E.~G. C.~P.}\
  \bibnamefont {van Loon}}, \bibinfo {author} {\bibfnamefont {H.}~\bibnamefont
  {Hafermann}}, \ and\ \bibinfo {author} {\bibfnamefont {M.~I.}\ \bibnamefont
  {Katsnelson}},\ }\href {\doibase 10.1103/PhysRevB.97.085125} {\bibfield
  {journal} {\bibinfo  {journal} {Phys. Rev. B}\ }\textbf {\bibinfo {volume}
  {97}},\ \bibinfo {pages} {085125} (\bibinfo {year}
  {2018}{\natexlab{b}})}\BibitemShut {NoStop}%
\bibitem [{\citenamefont {Guerci}\ \emph {et~al.}(2018)\citenamefont {Guerci},
  \citenamefont {Capone},\ and\ \citenamefont {Fabrizio}}]{Guerci18}%
  \BibitemOpen
  \bibfield  {author} {\bibinfo {author} {\bibfnamefont {D.}~\bibnamefont
  {Guerci}}, \bibinfo {author} {\bibfnamefont {M.}~\bibnamefont {Capone}}, \
  and\ \bibinfo {author} {\bibfnamefont {M.}~\bibnamefont {Fabrizio}},\
  }\href@noop {} {\enquote {\bibinfo {title} {Exciton mott transition
  revisited},}\ } (\bibinfo {year} {2018}),\ \Eprint
  {http://arxiv.org/abs/arXiv:1810.01843} {arXiv:1810.01843} \BibitemShut
  {NoStop}%
\bibitem [{\citenamefont {Krien}\ \emph {et~al.}(2018)\citenamefont {Krien},
  \citenamefont {van Loon}, \citenamefont {Katsnelson}, \citenamefont
  {Lichtenstein},\ and\ \citenamefont {Capone}}]{Krien18}%
  \BibitemOpen
  \bibfield  {author} {\bibinfo {author} {\bibfnamefont {F.}~\bibnamefont
  {Krien}}, \bibinfo {author} {\bibfnamefont {E.~G. C.~P.}\ \bibnamefont {van
  Loon}}, \bibinfo {author} {\bibfnamefont {M.~I.}\ \bibnamefont {Katsnelson}},
  \bibinfo {author} {\bibfnamefont {A.~I.}\ \bibnamefont {Lichtenstein}}, \
  and\ \bibinfo {author} {\bibfnamefont {M.}~\bibnamefont {Capone}},\
  }\href@noop {} {\enquote {\bibinfo {title} {Two-particle {F}ermi liquid
  parameters at the {M}ott transition},}\ } (\bibinfo {year} {2018}),\ \Eprint
  {http://arxiv.org/abs/arXiv:1811.00362} {arXiv:1811.00362} \BibitemShut
  {NoStop}%
\bibitem [{Note5()}]{Note5}%
  \BibitemOpen
  \bibinfo {note} {For the single-band Hubbard model $U^{\delimiter "3222378
  \delimiter "3222378 \delimiter "3223379 \delimiter "3223379 }=U^{\delimiter
  "3223379 \delimiter "3223379 \delimiter "3222378 \delimiter "3222378
  }=-U^{\delimiter "3223379 \delimiter "3222378 \delimiter "3222378 \delimiter
  "3223379 }=-U^{\delimiter "3222378 \delimiter "3223379 \delimiter "3223379
  \delimiter "3222378 }=U$, the other elements are zero.}\BibitemShut {Stop}%
\bibitem [{Note6()}]{Note6}%
  \BibitemOpen
  \bibinfo {note} {It is unclear whether a generalization of TPSC-DMF to
  symmetry-broken phases inherits thermodynamic consistency at second order
  critical points from the DMFT approximation~\cite {Krienthesis}, which may be
  clarified in future work.}\BibitemShut {Stop}%
\bibitem [{\citenamefont {Mahan}(2000)}]{Mahan00}%
  \BibitemOpen
  \bibfield  {author} {\bibinfo {author} {\bibfnamefont {G.~D.}\ \bibnamefont
  {Mahan}},\ }\href {\doibase 10.1007/978-1-4757-5714-9} {\emph {\bibinfo
  {title} {Many-Particle Physics}}}\ (\bibinfo  {publisher} {Springer {US}},\
  \bibinfo {year} {2000})\BibitemShut {NoStop}%
\bibitem [{\citenamefont {Krien}(2018)}]{Krienthesis}%
  \BibitemOpen
  \bibfield  {author} {\bibinfo {author} {\bibfnamefont {F.}~\bibnamefont
  {Krien}},\ }\href {http://ediss.sub.uni-hamburg.de/volltexte/2018/9182}
  {\enquote {\bibinfo {title} {Conserving dynamical mean-field approaches to
  strongly correlated systems},}\ } (\bibinfo {year} {2018})\BibitemShut
  {NoStop}%
\end{thebibliography}%

\end{document}